\documentclass[
  aps,
  prl,
  reprint,
  amsmath,
  amssymb,
  superscriptaddress,
  floatfix,
]{revtex4-2}

\usepackage[T1]{fontenc}
\usepackage{hyperref}
\usepackage{amsmath,bm,graphicx,multirow}
\usepackage{graphicx}
\usepackage{dcolumn}
\usepackage{bm}
\usepackage{float}
\usepackage{booktabs}
\usepackage{color}

    \makeatletter
\def\@fnsymbol#1{\ensuremath{\ifcase#1\or \dagger\or *\or \ddagger\or
   \mathsection\or \mathparagraph\or \|\or **\or \dagger\dagger
   \or \ddagger\ddagger \else\@ctrerr\fi}}
    \makeatother

\begin{document}
\let\suppAddContentsLine\addcontentsline
\renewcommand{\addcontentsline}[3]{}

\title{Evidence for itinerant electron-local moment interaction in Li-doped $\alpha$-MnTe}

\author{Tingjun\,Zhang}
\altaffiliation{These authors contributed equally to this work.}
\affiliation{Department of Physics and Astronomy, Rice University, Houston, Texas 77005, USA}
\affiliation{Rice Laboratory for Emergent Magnetic Materials and Smalley-Curl Institute, Rice University, Houston, Texas 77005, USA}
\affiliation{Applied Physics Graduate Program, Smalley-Curl Institute, Rice University, Houston, Texas 77005, USA}

\author{Steven\,J.\,Gomez\,Alvarado}
\altaffiliation{These authors contributed equally to this work.}
\affiliation{Department of Physics and Astronomy, Rice University, Houston, Texas 77005, USA}
\affiliation{Rice Laboratory for Emergent Magnetic Materials and Smalley-Curl Institute, Rice University, Houston, Texas 77005, USA}

\author{Sijie\,Xu}
\affiliation{Department of Physics and Astronomy, Rice University, Houston, Texas 77005, USA}
\affiliation{Rice Laboratory for Emergent Magnetic Materials and Smalley-Curl Institute, Rice University, Houston, Texas 77005, USA}

\author{Thomas\,A.\,Hulse}
\affiliation{Department of Physics and Astronomy, Rice University, Houston, Texas 77005, USA}
\affiliation{Applied Physics Graduate Program, Smalley-Curl Institute, Rice University, Houston, Texas 77005, USA}
\affiliation{Rice Laboratory for Emergent Magnetic Materials and Smalley-Curl Institute, Rice University, Houston, Texas 77005, USA}

\author{Yucheng\,Guo}
\affiliation{Department of Physics and Astronomy, Rice University, Houston, Texas 77005, USA}
\affiliation{Rice Laboratory for Emergent Magnetic Materials and Smalley-Curl Institute, Rice University, Houston, Texas 77005, USA}

\author{Shuting\,Peng}
\affiliation{Department of Physics, University of California, Berkeley, California 94720, USA}
\affiliation{Department of Physics and Astronomy, Rice University, Houston, Texas 77005, USA}

\author{Travis\,J.\,Williams}
\affiliation{ISIS Neutron and Muon Source, STFC Rutherford Appleton Laboratory, Didcot, Oxfordshire, OX11 0QX, United Kingdom}

\author{Xiaoping\,Wang}
\affiliation{Neutron Scattering Division, Oak Ridge National Laboratory, Oak Ridge, Tennessee 37831, USA}

\author{Junhong\,He}
\affiliation{Neutron Scattering Division, Oak Ridge National Laboratory, Oak Ridge, Tennessee 37831, USA}

\author{Matthew\,B.\,Stone}
\affiliation{Neutron Scattering Division, Oak Ridge National Laboratory, Oak Ridge, Tennessee 37831, USA}

\author{Colin\,Sarkis}
\affiliation{Neutron Scattering Division, Oak Ridge National Laboratory, Oak Ridge, Tennessee 37831, USA}

\author{Feng\,Ye}
\affiliation{Neutron Scattering Division, Oak Ridge National Laboratory, Oak Ridge, Tennessee 37831, USA}

\author{Aaron\,Bostwick}
\affiliation{Advanced Light Source, Lawrence Berkeley National Laboratory, Berkeley, California 94720, USA}

\author{Christopher\,Jozwiak}
\affiliation{Advanced Light Source, Lawrence Berkeley National Laboratory, Berkeley, California 94720, USA}

\author{Eli\,Rotenberg}
\affiliation{Advanced Light Source, Lawrence Berkeley National Laboratory, Berkeley, California 94720, USA}

\author{Zhaoyu\,Liu}
\affiliation{Department of Physics and Astronomy, Rice University, Houston, Texas 77005, USA}
\affiliation{Rice Laboratory for Emergent Magnetic Materials and Smalley-Curl Institute, Rice University, Houston, Texas 77005, USA}

\author{Jinyulin\,Li}
\affiliation{Department of Physics and Texas Center for Superconductivity at the University of Houston, Houston, Texas 77204, USA}

\author{Aparna\,Jayakumar}
\affiliation{Department of Physics and Astronomy, Rice University, Houston, Texas 77005, USA}
\affiliation{Rice Laboratory for Emergent Magnetic Materials and Smalley-Curl Institute, Rice University, Houston, Texas 77005, USA}

\author{Zehao\,Wang}
\affiliation{Department of Physics and Astronomy, Rice University, Houston, Texas 77005, USA}
\affiliation{Rice Laboratory for Emergent Magnetic Materials and Smalley-Curl Institute, Rice University, Houston, Texas 77005, USA}

\author{Yaofeng\,Xie}
\affiliation{Department of Physics and Astronomy, Rice University, Houston, Texas 77005, USA}
\affiliation{Rice Laboratory for Emergent Magnetic Materials and Smalley-Curl Institute, Rice University, Houston, Texas 77005, USA}

\author{Yiheng\,Wang}
\affiliation{Department of Physics and Astronomy, Rice University, Houston, Texas 77005, USA}
\affiliation{Rice Laboratory for Emergent Magnetic Materials and Smalley-Curl Institute, Rice University, Houston, Texas 77005, USA}

\author{Ching-Wu\,Chu}
\affiliation{Department of Physics and Texas Center for Superconductivity at the University of Houston, Houston, Texas 77204, USA}

\author{Liangzi\,Deng}
\affiliation{Department of Physics and Texas Center for Superconductivity at the University of Houston, Houston, Texas 77204, USA}

\author{Emilia\,Morosan}
\affiliation{Department of Physics and Astronomy, Rice University, Houston, Texas 77005, USA}
\affiliation{Rice Laboratory for Emergent Magnetic Materials and Smalley-Curl Institute, Rice University, Houston, Texas 77005, USA}

\author{Robert\,J.\,Birgeneau}
\affiliation{Department of Physics, University of California, Berkeley, California 94720, USA}
\affiliation{Materials Science Division, Lawrence Berkeley National Laboratory, Berkeley, California 94720, USA}

\author{Ming\,Yi}
\affiliation{Department of Physics and Astronomy, Rice University, Houston, Texas 77005, USA}
\affiliation{Rice Laboratory for Emergent Magnetic Materials and Smalley-Curl Institute, Rice University, Houston, Texas 77005, USA}

\author{Pengcheng\,Dai}
\email{pdai@rice.edu}
\affiliation{Department of Physics and Astronomy, Rice University, Houston, Texas 77005, USA}
\affiliation{Rice Laboratory for Emergent Magnetic Materials and Smalley-Curl Institute, Rice University, Houston, Texas 77005, USA}

\begin{abstract}

{\color{black}
We use inelastic neutron scattering (INS) and angle-resolved photoemission spectroscopy (ARPES) to study the impact of Li doping on the semiconducting altermagnet $\alpha$-MnTe.
Introducing Li reorients the ordered moment direction from in-plane to out-of-plane and increases the density of itinerant carriers.
While our ARPES measurements do not indicate any notable doping-induced changes in the electronic band structure or the magnitude of the altermagnetic band splitting, our INS measurements reveal an abrupt doping-induced decrease in the spin-wave lifetime near the zone boundary at high energies. This finding is consistent with a new magnon decay channel driven by subtle, doping-induced changes in the band structure and enhanced interactions between Mn$^{2+}$ local moments and itinerant electrons.
By extracting the local dynamic susceptibility from INS spectra and applying the total moment sum rule, we find that both undoped and Li-doped $\alpha$-MnTe exhibit a nearly full expected Mn$^{2+}$ local moment of $\approx5.9~\mu_\mathrm{B}$ with $S=5/2$. These findings suggest that $\alpha$-MnTe hosts robust local-moment altermagnetism which shows a breakdown at high energies upon addition of itinerant carriers, highlighting the importance of carrier-spin coupling in magneto-transport and spin dynamics of altermagnets, even in the dilute-carrier limit.}
\end{abstract}

\maketitle

\noindent
{\color{black}In conventional local-moment (anti)ferromagnets governed by Heisenberg exchange interactions, collective excitations of localized moments in the magnetically ordered state manifest as} spin waves (magnons) characterized by linear spin wave theories (LSWT) that ignore all terms of order higher than quadratic \cite{WOS:000207403700002,PhysRev.58.1098,PhysRev.117.117,10.1093/oso/9780198862314.001.0001}.
Their dispersion determines the nearest-neighbor (NN), next-nearest-neighbor (NNN), etc. exchange couplings, and, therefore, the magnetic ordering temperature --- {\color{black}largely independent} of the ordered moment direction controlled by spin-orbit coupling (SOC)
\cite{10.1093/oso/9780198862314.001.0001}.
{\color{black} In simple, collinear magnets, higher-order} magnon interactions are often weak \cite{RevModPhys.85.219}, and thus the magnon lifetime as measured by the energy width of spin wave spectra should be {\color{black} effectively} infinite and only limited by the instrumental resolution as seen in ferromagnetic EuO \cite{PhysRevB.14.4923} and collinear antiferromagnetic Rb$_2$MnF$_4$ \cite{PhysRevLett.111.017204}.
For classical spin systems (\textit{i.e.},$\ S \gg 1/2$) without strong quantum fluctuations, geometric frustration, or SOC,
magnon decay processes are generally not expected in insulating magnets \cite{RevModPhys.85.219}.

\begin{figure}[t]
    \includegraphics[width=0.9\linewidth]{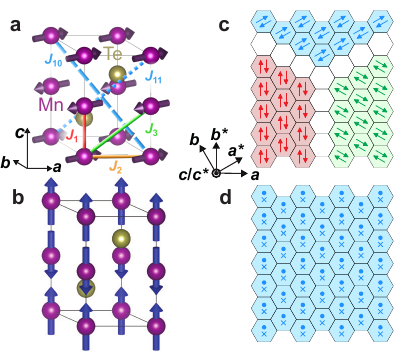}
    \caption{\label{fig:fig1}
    (a,b) Parallel view of the crystal and magnetic structures of MnTe and Li-MnTe, respectively, highlighting magnetic moment directions and exchange couplings. Dashed lines indicate the altermagnetic exchange interactions $J_{10}$ and $J_{11}$.
    (c,d) Schematics of potential magnetic domain configurations MnTe and Li-MnTe, respectively.
    }
\end{figure}

{\color{black}
In metallic magnets, magnetic order can arise either from Fermi-surface instabilities of itinerant electrons \cite{RevModPhys.60.209} or from interactions between preformed localized moments.
The associated spin waves can thus manifest either as coherent, collective bound states of spin-flip particle-hole pairs or as conventional transverse excitations of localized spins.
In either regime, these spin waves can decay by transferring their energy and angular momentum into the continuum of incoherent spin-flip particle-hole excitations, a damping process governed sensitively by the electronic band structure \cite{Stoner1936,Stoner1938,PhysRev.52.198,RevModPhys.25.211}.
In strongly itinerant {\color{black}ferromagnets}, magnons may enter the Stoner particle-hole continuum at elevated energies, where their lifetime is rapidly reduced via Landau damping \cite{Stoner1936,Stoner1938,PhysRev.52.198,RevModPhys.25.211}.  Spin waves in elemental $\alpha$-Fe, for example, are observed to remain essentially resolution-limited up to energies approaching the onset of the Stoner continuum near 200~meV \cite{PhysRevB.11.2624,10.1063/1.348814,brookes_spin_2020}.
For compensated antiferromagnets, the damping behavior is expected to be qualitatively different as spin-degenerate electronic bands may permit intraband electron spin-flip scattering without the energy threshold associated with exchange-split bands in ferromagnets \cite{simensen_magnon_2020}. At long wavelengths and low energies, the resulting electron-mediated relaxation can be described phenomenologically within a generalized Landau--Lifshitz--Gilbert framework \cite{Landau1935,Gilbert1955}, while the damping of finite-wave-vector spin waves depends on the momentum-dependent phase space and matrix elements for electron--hole excitations. Consequently, measurable spin-wave broadening in metallic magnets provides a sensitive probe of the degree of itinerancy and of the details of carrier-spin coupling encoded in the electronic structure near the Fermi level.}

In the semiconducting altermagnet $\alpha$-MnTe {\color{black}(hereafter, MnTe)}, opposite magnetic sublattices of Mn$^{2+}$ are connected by symmetry operations that are not inversion ($\mathcal{P}$) or translation ($t$) symmetric due to the specific local crystal environment [Fig. \ref{fig:fig1}(a)]. This leads to alternating spin splitting of electronic energy bands, arising from exchange coupling rather than SOC \cite{smejkal_CrystalTimereversalSymmetry_2020, smejkal_ConventionalFerromagnetismAntiferromagnetism_2022, smejkal_EmergingResearchLandscape_2022}, in contrast to collinear antiferromagnets, where the combination of time-reversal symmetry ($\mathcal{T}$) and $\mathcal{P}$ or $t$ symmetries between opposite magnetic sublattices enforces their degeneracy in the limit of zero SOC \cite{10.1093/oso/9780198862314.001.0001}. Indeed, several phenomena commonly associated with the breaking of $\mathcal{T}$-symmetry --- spin-split electronic bands \cite{din_unconventional_2025}, anomalous Hall effects (AHE) \cite{gonzalez_betancourt_spontaneous_2023, kluczyk_coexistence_2024, PhysRevLett.134.086701,liu_StraintunableAnomalousHall_2025, smolenski2025straintunabilitymultipolarberrycurvature},
and magnetic circular dichroism \cite{hariki_x-ray_2024} --- have been experimentally realized in altermagnetic materials \cite{barman_2021_2021,flebus_2024_2024}.
{\color{black}Altermagnetic symmetries may also give rise to a momentum-dependent energy splitting of chiral magnons, which is non-relativistic in origin and holds even in the limit of symmetric exchange interactions \cite{liu_chiral_2024,CrSb_alter,Fe2O3_alter,FeF2_chiral,MnF2_chiral,liu_pol_INS}, as well as a finite nuclear--magnetic interference term which distinguishes altermagnetism from conventional antiferromagnetism \cite{PhysRevB.111.L060405,liu_pol_diff}. At the microscopic level, the chiral splitting of magnons in MnTe originates from the inequivalent exchange pathways $J_{10}$ and $J_{11}$ which, despite having equal distance, are distinguished by their distinct superexchange routes through Te orbitals [Fig.~\ref{fig:fig1}(a)] \cite{liu_chiral_2024}.}

An attractive aspect of MnTe in the context of spintronics is the tunability of its altermagnetic ground state, which can be modified by
chemical substitution \cite{moseley_giant_2022, yumnam_MagnonGapTuning_2024},
uniaxial strain \cite{PhysRevLett.134.086701,liu_StraintunableAnomalousHall_2025,smolenski2025straintunabilitymultipolarberrycurvature},
and external applied fields \cite{kriegner_multiple-stable_2016, kriegner_magnetic_2017, amin_nanoscale_2024, gonzalez_betancourt_anisotropic_2024, dzian_antiferromagnetic_2025}. MnTe adopts a layered $A$-type magnetic order below $T_\mathrm{AM}=307$~K \cite{kunitomiNeutronDiffractionStudy1964,efremdsaLowtemperatureNeutronDiffraction2005,szuszkiewiczNeutronScatteringStudy2005,kriegnerMagneticAnisotropyAntiferromagnetic2017,szuszkiewiczSpinwaveMeasurementsHexagonal2006}, {\color{black} with in-plane Mn$^{2+}$ moments oriented in the basal plane [Fig. \ref{fig:fig1}(a,c)], while the precise ground-state moment direction remains under debate \cite{liu_StraintunableAnomalousHall_2025, liebman-pelaez_strain_2026}.
In the absence of a symmetry-breaking field, multiple in-plane magnetic domains
result in a magnetic response which is domain-averaged [Fig. \ref{fig:fig1}(c)].
Less than 1\% Li doping modifies the anisotropy of these Mn$^{2+}$ moments from in-plane to out-of-plane and introduces charge carriers into the system [Fig. \ref{fig:fig1}(b,d)] \cite{yumnam_MagnonGapTuning_2024}.
This removes any in-plane domain-averaging in the magnetic sector without modifying the symmetries required for altermagnetism [Fig. \ref{fig:fig1}(d)] \cite{smejkal_CrystalTimereversalSymmetry_2020, smejkal_ConventionalFerromagnetismAntiferromagnetism_2022, smejkal_EmergingResearchLandscape_2022}.}

\begin{figure*}[t]
    \includegraphics[width=0.95\linewidth]{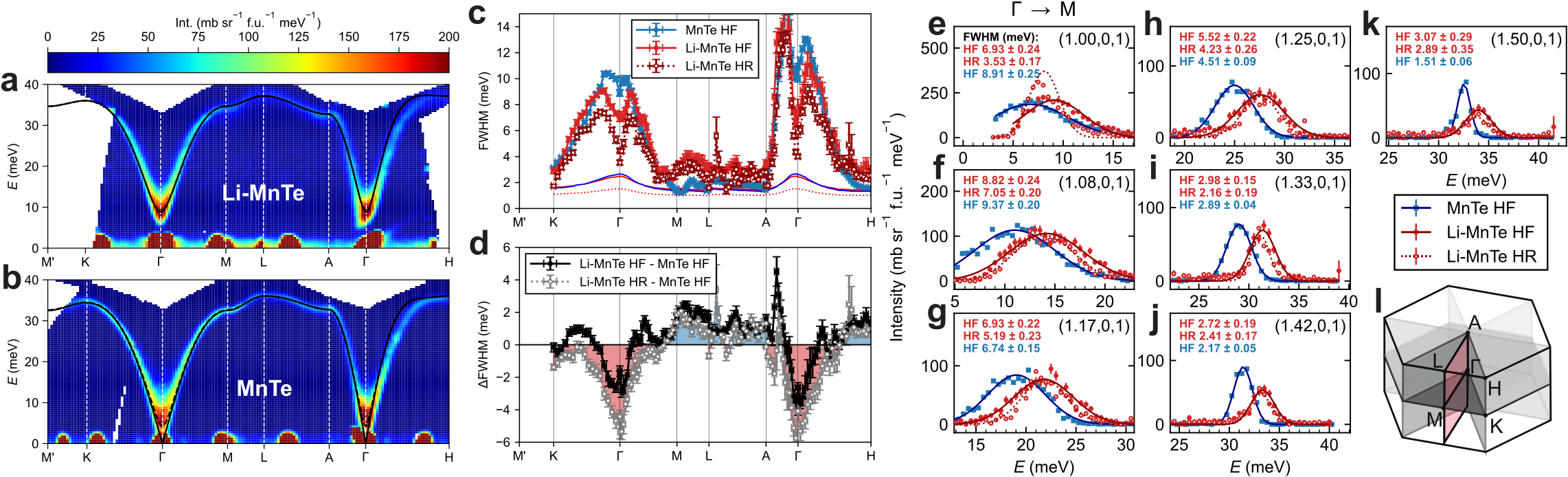}

    \caption{\label{fig:fig2}(a,b) INS spectra acquired at base temperature ($T\approx5$ K) in HF mode with an incident energy $E_i = 50$ meV for Li-MnTe and MnTe, shown along high-symmetry directions within the nodal planes.
    Black curves represent calculated spin wave branches.
    {\color{black}
    (c) Energy FWHM obtained from Gaussian fits within the nodal planes for pure and Li-MnTe. Solid and dashed curves represent energy-dependent instrumental resolutions (Fig. S1).
    (d) $\Delta$FWHM obtained by subtracting the FWHM of pristine MnTe from that of Li-MnTe. Vertical error bars represent 1$\sigma$ standard deviations and horizontal error bars indicate the $\bm{Q}$ integration range.
    (e-k) Energy-dependent line cuts at select $\bm{Q}$ points along the $\Gamma$-M path. Fitted FWHM values and uncertainties are listed within each panel.
    Filled symbols use Li-MnTe HF data and open symbols use Li-MnTe HR data in each panel.
    (l) Schematic of the BZ, where nodal planes are shaded in gray and the $\Gamma$-A--L--M plane in red.
    }
    }
\end{figure*}

Understanding the influence of charge carriers on the prototypical altermagnetic ground state in MnTe and its collective excitations is important. {\color{black}The momentum-dependent spin splitting of the electronic bands may introduce additional complexity into the available electron--magnon scattering channels, resulting in anisotropic and energy-/momentum-dependent magnon damping \cite{smejkal_EmergingResearchLandscape_2022,costa2025giant}.} Further, the AHE appears to depend sensitively on sample-dependent variations in the charge gap, indicating that a narrow window of in-gap impurity states is required for its observation in MnTe \cite{liu_StraintunableAnomalousHall_2025}.

In Li-doped $\alpha$-MnTe (hereafter, Li-MnTe), the extent to which doped charge carriers and the associated substitutional disorder affect the spin dynamics underpinning the altermagnetic functionality remains unresolved. To address this, we investigate the effects of $3\%$ Li doping on the magnetic excitations and electronic structure of Li-MnTe using inelastic neutron scattering (INS) and angle-resolved photoemission spectroscopy (ARPES), respectively. We find that, while the altermagnetic electronic spin splitting remains essentially unmodified, Li doping produces several notable changes in the spin dynamics including a pronounced magnon damping near the zone boundary, pointing to enhanced interactions between itinerant electrons and Mn$^{2+}$ local moments which is essential for understanding the emergence of the AHE \cite{liu_StraintunableAnomalousHall_2025} as well as the spin dynamics in this system.

{\color{black} While our pristine MnTe single crystal is a weakly $p$-type, self-doped semiconductor, our Li-MnTe single crystal exhibits a low-resistivity metallic state domainated by hole-like carriers [Fig. \ref{fig:fig5}(b,d)], consistent with prior reports on polycrystalline samples \cite{ren_synergistic_2017,zheng_paramagnon_2019}.
The carrier concentration extracted from our Hall-effect measurements on Li-MnTe suggest approximately 3\% excess holes per formula unit generated by Li doping. We attempted to determine where Li incorporates into the lattice by refining several structural models against single crystal neutron diffraction data collected on the TOPAZ diffractometer at the Spallation Neutron Source (SNS). However, we find that placing Li on the Mn $2a$ site, the interstitial $2d$ and $4f$ sites, and the Te $2c$ site yields nearly identical fit metrics, and thus we cannot conclusively establish the location of Li within the structure.
Previous studies have inferred alkali-metal (Li$^{+}$, Na$^{+}$) doping on the Mn$^{2+}$ $2a$ site consistent with the enhanced $p$-type electronic character in polycrystalline samples \cite{ren_synergistic_2017,zheng_paramagnon_2019}, while recent neutron powder diffraction results place Li on the interstitial $2d$ site \cite{moseley_giant_2022}. Our refinements also suggest a Te deficiency of approximately 4\% in both samples, though this would be inconsistent with the observed $p$-type character in the absence of compensating defects \cite{zulkifal_multiple_2023}, and Te vacancies are expected to be thermodynamically unfavorable during crystal growth in a Te flux \cite{huynh_intrinsic_2026}.

}

To understand the impact of Li-doping on the spin dynamics,
we conducted time-of-flight INS measurements on the SEQUOIA spectrometer at SNS for pristine MnTe and on the MAPS Spectrometer at the ISIS Neutron Source for Li-MnTe. Both measurements used a single crystal with mosaic less than 1 degree, using nearly identical instrumental resolutions (Fig.~S1) and normalized to absolute units \cite{10.1063/1.4818323}.
In pristine MnTe, while spin waves of opposite chirality remain degenerate in the nodal planes shaded in gray in {Fig. \ref{fig:fig2}f}, this degeneracy is lifted in the $\Gamma$-A-L-M plane highlighted in red \cite{liu_chiral_2024}.
We first examine the nodal planes [Fig. \ref{fig:fig2} (a,b)], which reveal three differences induced by Li doping:
(1) the opening of a $\sim$6~meV spin gap indicative of the change in single-ion anisotropy,
(2) a reduction in the overall bandwidth from 35~meV to 32~meV, signaling a 9\% reduction in the effective exchange energy scale, and
(3) a notably broadened linewidth near the zone boundary [Fig. \ref{fig:fig2}(c-e)].
{\color{black} The bandwidth reduction in (2) could reflect a carrier-mediated modification of the exchange interactions, for example through an Ruderman-Kittel-Kasuya-Yosida (RKKY)-type contribution that competes with the dominant antiferromagnetic exchange. The magnon damping implied by (3) may have several possible physical origins in metallic magnets, as previously discussed.}

To distinguish between these {\color{black}possible origins}, we compare the energy full width at half-maximum (FWHM) using constant-$\bm{Q}$ line cuts in high-flux (HF) mode, where instrumental resolutions are nearly identical (Fig. S1). We also carried out high-resolution (HR) measurements for Li-MnTe in order to differentiate between intrinsic and extrinsic broadening effects (Figs. S3, S4). {\color{black} While spin waves in MnTe near $\Gamma$ are split into two branches due to single-ion anisotropy [Figs. \ref{fig:fig6}(a,c)], the corresponding branches in Li-MnTe are degenerate [Figs. \ref{fig:fig6}(b) and (d)]. Thus, the apparent linewidth near the zone center appears unusually broad [Figs. \ref{fig:fig2}(c,d)]. We therefore compare the FWHM slightly away from $\Gamma$, where the influence of mode overlap is reduced.} Taking the MnTe FWHM to represent the intrinsic linewidth, the difference profile ($\Delta$FWHM) between the two HF datasets demonstrates that the linewidth is nearly identical between the two samples away from the BZ boundary [Fig. \ref{fig:fig2}(d)], with the exception of an artifact at intermediate $\bm{Q}$ (\textit{e.g.}, halfway along $\Gamma$-M) which manifests as a hump in the HF dataset but is less pronounced in the HR dataset. In contrast, near the zone boundary we find {\color{black}an abrupt and} substantial increase in the FWHM of Li-MnTe in both HF and HR modes, strongly suggesting that this is an intrinsic effect induced by Li doping.

\begin{figure}[t]
    \includegraphics[width=0.95\linewidth]{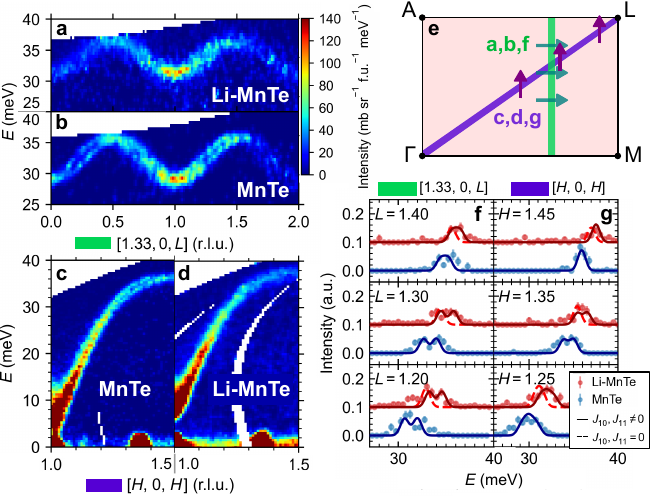}

    \caption{\label{fig:fig3}(a,b) INS spectra along $[1.33,0,L]$ and (c,d) along the $[H,0,H]$ direction for MnTe and Li-MnTe.
    (e) Schematic of the $\Gamma$-A-L-M plane of the BZ, where splitting is expected to occur in MnTe.
    Green and purple lines highlight the $[1.33,0,L]$ and $[H,0,H]$ directions, respectively.
    Arrows indicate the direction of constant-$\bm{Q}$ line cuts.
    (f) Comparison of line cuts for the dispersion along the $[1.33,0,L]$ and
    (g) $[H,0,H]$ directions.
    Points represent the measured data, and error bars represent 1$\sigma$ uncertainties.
    Solid and dashed curves represent fits to LSWT models where $J_{10}$ and $J_{11}$ are nonzero and zero, respectively.
    For Li-MnTe, the nonzero values of $J_{10}$ and $J_{11}$ were fixed to the refined values from MnTe.
    Data for Li-MnTe was shifted vertically by 0.075 units for clarity.
    }
\end{figure}

{\color{black}

The abrupt, anisotropic broadening is inconsistent with the gradual onset expected for damping dominated by disorder effects. In general, random disorder may manifest as variations in the local exchange couplings by interstitial Li or as nonmagnetic ``vacancies'' within the ordered magnetic state by substitutional Li on the Mn $2a$ site.
For three-dimensional disordered magnets in the dilute-defect limit, the disorder-averaged magnon self-energy scales with defect concentration $x$ according to the single-impurity $T$-matrix \cite{Izyumov1966,Brenig1991} and non-resonant scattering from either random exchange bonds or nonmagnetic vacancies is expected to produce a slowly varying evolution of the linewidth across the BZ \cite{Chernyshev2002,Buczek2016,hameed_effect_2023}.
Alternatively, resonant scattering from impurity-induced quasilocalized states may generate sharper, energy-dependent features at an intermediate resonance energy \cite{Izyumov1966,Brenig1991,Chernyshev2002} as observed in magnetically diluted altermagnet $\text{Mn}_{1-x}\text{Zn}_x\text{F}_2$ \cite{svensson_resonant_1969}, though the associated linewidth broadening is expected to return to the original linewidth above the resonance.
Instead, the broadening in Li-MnTe has an onset energy which appears to have an anisotropic momentum dependence and varies by a few meV depending on $\bm{Q}$ (Fig. S5), while the linewidth at the top of the dispersion remains far from resolution-limited [Fig. 2(c,d)], suggesting that distinct decay processes must be considered.
This behavior also differs qualitatively from the damping attributed to dynamical spin-lattice coupling in ferromagnet CrGeTe$_3$, which also produces a gradual (quadratic) energy linewidth increase with $\bm{Q}$ \cite{WOS:000895384200001}.
A related finite linewidth broadening has been reported in low-carrier-density Mn-based metallic magnets such as YbMnBi$_2$ \cite{sapkota_signatures_2020} and YbMnSb$_2$ \cite{hu_coupling_2023}, though in these cases the observed broadening is uniform across the BZ.
}

As a quantitative measure of the modifications to the spin Hamiltonian within a local-moment framework, we fit the spin wave spectra of both compounds using LSWT based on a Heisenberg Hamiltonian with single-ion anisotropy ($D_z$) (Eqn. S1). A positive (negative) $D_z$ corresponds to easy-plane (easy-axis) anisotropy in the undoped (doped) system.
Compared to MnTe, the NN exchange $J_1$ in Li-MnTe is reduced by $\sim$8\%, consistent with the spin wave bandwidth reduction.
We modeled the measured energy dependence of the spectral weight at each momentum transfer {\color{black}[$I(E)$]} as a Gaussian profile with a finite HWHM$=$FWHM/2.
We then constructed three independent LSWT models for Li-MnTe based on distinct input dispersions along the BZ-boundary segments M-L and L-A, where the linewidth broadening is most pronounced: (1) a ``center'' model, employing the dispersion defined by the peak energies $E(\bm{Q})$ obtained directly from Gaussian fits to the experimental spectra, (2) a ``lower'' model, employing energies shifted downward via $E(\bm{Q})-\mathrm{HWHM}(\bm{Q})$, and (3) an ``upper'' model, employing energies shifted upward via $E(\bm{Q})+\mathrm{HWHM}(\bm{Q})$ (Table S1).
Among the exchange parameters, the further-neighbor interactions and the anisotropy term were found to be relatively insensitive to this spectral weight distribution.
{\color{black}In contrast, the $J_1$ term exhibited a distribution range of $\Delta J_1\approx 0.33$ meV, corresponding to a HWHM of about 1.5~meV at the BZ boundary, or equivalently $\Delta J_1/J_1 \approx 0.10$.
However, a superposition of the spin wave spectra obtained from these models remains incapable of capturing the abrupt onset of broadening observed in our INS spectra (Fig. S2), disfavoring a simple distribution of exchange couplings.
}

\begin{figure}
    \includegraphics[width=0.95\linewidth]{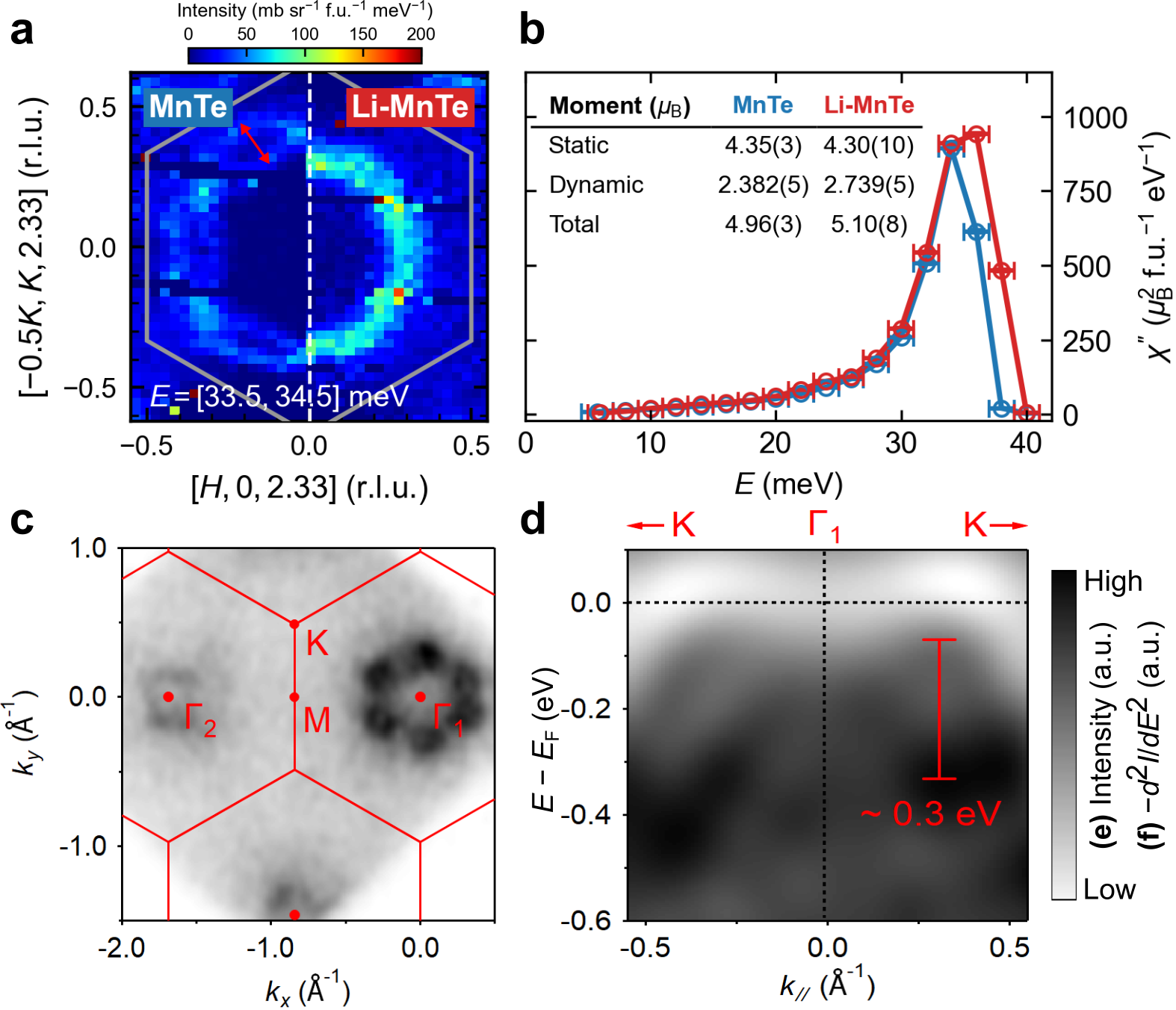}

    \caption{\label{fig:fig4}(a) Constant-energy map of the INS spectra in the $HK$ plane for pure (left) and Li-MnTe (right) at $E=34$~meV. The in-plane BZ is outlined by a thin grey line. The observed splitting in pure MnTe is indicated by red arrows. (b) Energy dependence of the local dynamic susceptibility extracted from INS data. Inset table describes static and dynamic moments obtained from neutron diffraction and INS data, respectively. (c) ARPES 100~meV integrated constant-energy contour of Li-MnTe centered around the Fermi level, showing the first and second electronic BZ. (d) Second-energy-derivative high-symmetry cut along K-$\Gamma_1$-K with altermagnetic splitting of $\approx300$~meV indicated by red bars.
    }
\end{figure}

We now inspect the spin wave spectra on the $\Gamma$-A-L-M plane of the BZ, where splitting is expected in MnTe [Fig. \ref{fig:fig3}(e)] \cite{liu_chiral_2024}. Figure \ref{fig:fig3}(a-d) compares spin wave spectra along $[1.33,0,L]$ and $[H,0,H]$ directions for both systems.
While splitting is clearly observed in MnTe \cite{liu_chiral_2024}, the broadened linewidth in Li-MnTe makes the splitting more challenging to resolve.
This is also reflected in constant energy cuts [Fig. \ref{fig:fig4}(a)], where inspection of the $\bm{Q}$-dependence of the ring-like spectral weight {\color{black}reveals six intense nodes marking the nodal planes where the two chiral branches recombine in MnTe.} We note that in Li-MnTe a weak alternation is evident at similar energies, which is consistent with a persistent splitting [Fig. \ref{fig:fig4}(a)].
To more closely investigate the presence of magnon splitting in the two systems, we performed constant-$\bm{Q}$ cuts near the expected splitting position in $\bm{Q}$-$E$ space [Fig. \ref{fig:fig3}(e-g)].
While two peaks are resolvable in the line cuts for MnTe, the spectra for Li-MnTe are broader and only a single, broad feature is resolvable for most $L$ values.

To better understand whether the observed dispersion and lineshape for Li-MnTe can be explained by a persistent splitting which is simply broadened by the addition of itinerant carriers, we fit the data to two distinct LSWT models with refined $J_1$,\,$J_2$,\,$J_3$: (1) with $J_{10}$ and $J_{11}$ zero, where splitting is necessarily absent; (2) with non-zero difference $J_{10}-J_{11}$ fixed at the value refined for MnTe and $J_{10}$, $J_{11}$ allowed to refine under this constraint [Fig. \ref{fig:fig3}(f,g)]. The former scenario, identical to the ``center'' model presented earlier, underestimates both the energy and the linewidth near the expected splitting (Table S1). For the model with $J_{10},J_{11}\neq0$, we find improved agreement with experimental energies, further suggesting that the splitting in MnTe may survive Li-doping.

While Li-doping strongly modifies the spin anisotropy and tunes the carrier concentration, altermagnetism is, \textit{a priori}, expected to remain robust as its framework holds valid for both insulating and metallic systems.
{\color{black}To verify this, we performed ARPES measurements on Li-MnTe to interrogate the behavior of the electronic band splitting in the presence of Li dopants. As with MnTe, the Fermi surface at $k_z=0$ consists of six petal-like pockets around $\Gamma$ [Fig. \ref{fig:fig4}(c)] \cite{krempasky_altermagnetic_2024, osumi_observation_2024}. Along $\Gamma_1$-K, where altermagnetic splitting from SOC has previously been predicted and observed in MnTe \cite{krempasky_altermagnetic_2024, osumi_observation_2024, din_unconventional_2025}, we find splitting of approximately the same magnitude ($\approx$300~meV), verifying the robust nature of altermagnetism in 3\% Li-doped MnTe [Fig. \ref{fig:fig4}(d)].  Within the scope of these experiments, no obvious changes in the band structure near the Fermi level were observed when compared to MnTe. We note that previous calculations have predicted modifications to the band structure due to both light Li-doping \cite{moseley_giant_2022} and spin reorientation \cite{faria_junior_sensitivity_2023}. While our data clearly resolve the persistent altermagnetic spin splitting, they do not exclude the possibilities of subtle band modifications as the current data quality is constrained by well-known complications in MnTe related to poor in-situ cleaving. Further investigations of changes to the optical gap or ARPES spectrum may be able to resolve this. Thus, the observed magnon damping must arise from subtle changes in the electronic structure that impact the nature of the magnetic exchange and spin dynamics via coupling between itinerant electrons and local moments through RKKY (indirect exchange) interactions and magnon-electron scattering processes, respectively.

The fact that small amounts of Li doping can dramatically affect the spin gap and spin wave bandwidth is also supportive of the important role of itinerant electrons in this system.
Further, the sensitive dependence of the AHE on the charge gap in undoped MnTe \cite{liu_StraintunableAnomalousHall_2025} underscores the notion that the physics of this system can largely be described as a network of local moments that interact with itinerant charge carriers. We further test this notion by calculating the local dynamic susceptibility from the total spectral weight of our INS measurements [Fig. \ref{fig:fig4}(b)], which, when integrated across all energy transfers, yields a quantitative measure of the local dynamic moment \cite{10.1063/1.4818323,RevModPhys.87.855}. {\color{black}Our analysis yields a dynamic moment of 2.382(5)~$\mu_\mathrm{B}$ [2.739(5)~$\mu_\mathrm{B}$] for pristine MnTe~[Li-MnTe]. When combined with the corresponding ordered moment (Eqn. S2), this indicates a total effective moment of 4.96(3)~$\mu_\mathrm{B}$ [5.10(8)~$\mu_\mathrm{B}$] --- effectively the full $\mu_\mathrm{eff}=5.9~\mu_\mathrm{B}$ for $S=5/2$ ($\gg S=1/2$) Mn$^{2+}$ ions within the accuracy of this technique. This is consistent with (a) our expectation that the small but finite magnon lifetime broadening does not remove spectral weight and (b) it does not appear to push spectral weight outside our measured energy range, and thus Li doping largely leaves the Mn$^{2+}$ local-moment character intact.}

{\color{black}
In summary, {\color{black}anisotropic magnon damping is observed in Li-MnTe across the BZ boundary}, suggesting a mechanism based in subtle electronic structure changes with characteristic momentum and energy dependence distinct from the spin splitting of electronic bands observed via ARPES.
Additional chirality-dependent magnon decay processes have been proposed in altermagnets \cite{costa2025giant,cichutek2025spontaneous,eto2025spontaneous}, although whether these mechanisms can be directly observed by INS remains an open question.
This finding underscores the important role of interactions between itinerant carriers and local moments which must be considered to understand the physics of altermagnetic candidate materials and their potential for applications in next-generation spintronic devices.
}

\newpage
We thank Kirill Belashchenko, Erick Lawrence, and Yichen Zhang for helpful discussions.
The single-crystal synthesis, transport, neutron scattering, and ARPES work at Rice were supported by the U.S. DOE, BES under Grant Nos. DE-SC0026179 (P.D., M.Y., and E.M.). Part of the materials characterization efforts at Rice is supported by the Robert A. Welch Foundation Grant No. C-1839 (P.D.). X.W. acknowledges research sponsored by the LDRD Program of ORNL, managed by UT-Battelle, LLC, for the U.S. DOE.  Part of the materials characterization efforts at UH is supported by the T. L. L. Temple Foundation; the John J and Rebecca Moores Endowment; the State of Texas through the Texas Center for Superconductivity at UH; and the Robert A. Welch Foundation (00730-5021-H0452-B0001-G0512489). This research was supported in part by an appointment to the ORNL GRO Program, sponsored by the U.S. DOE and administered by the ORISE.
A portion of this research used resources at the Spallation Neutron Source, a DOE Office of Science User Facility operated by ORNL. The beam times were allocated to TOPAZ and SEQUOIA on Proposals No. IPTS-35052, IPTS-33830 and IPTS-34120. The experiment at the ISIS Neutron and Muon Source was supported by beam time allocations RB2510033 and RB2510079 from the Science and Technology Facilities Council. The ARPES work at Rice University was supported in part by the Gordon and Betty Moore Foundation's EPiQS Initiative through grant No. GBMF9470 and the Robert A. Welch Foundation Grant No. C-2175 (M.Y.). This research used resources of the Advanced Light Source, which is a DOE Office of Science User Facility under contract no. DE-AC02-05CH11231. Work at UCB and LBL was funded by the U.S. DOE, BES under Contract No. DE-AC02-05CH11231 (Quantum Materials Program KC2202).

The data that support the findings of this manuscript are openly available \cite{zenodo}.

\bibliographystyle{apsrev4-2}
\bibliography{biblio_cleaned}

\clearpage

\section*{End Matter}
{\color{black}
\textit{Transport and magnetization data}---
To further corroborate the magnetic moment orientations determined from neutron diffraction, we performed magnetic susceptibility measurements.
The magnetization of 3\% Li-doped MnTe is consistent with a N\'eel vector aligned along the $c$ axis [Fig. \ref{fig:fig5}(a)] \cite{yumnam_MagnonGapTuning_2024}, whereas the N\'eel vector in pure MnTe lies within the $ab$ plane.
The resistivity of Li-MnTe is reduced by approximately two orders of magnitude compared with that of pure MnTe [Fig. \ref{fig:fig5}(b)].
Notably, $T_\mathrm{AM}$ remains nearly unchanged at this doping level, reminiscent of the robust antiferromagnetic ground state observed in K-doped BaMn$_2$As$_2$ \cite{lamsal_persistence_2013}.
This suggests that the dominant exchange energy scale between Mn$^{2+}$ local moments is largely preserved upon Li doping.
The positive Hall coefficients indicate that holes are the dominant charge carriers in both compounds, while the extracted carrier densities and mobilities suggest that Li doping enhances the hole contribution to electrical conduction [Figs. \ref{fig:fig5}(c) and (d)].

\textit{LSWT calculation of spin wave dispersion}---
We performed LSWT calculations using previously reported exchange parameters for the Heisenberg Hamiltonian to simulate the spin wave dispersions of pure and Li-doped MnTe.
These calculations illustrate the contributions of different magnetic domains to the neutron scattering intensity and distinguish the magnon splitting induced by single-ion anisotropy from that arising from altermagnetic exchange interactions.
The spin wave dispersions from different magnetic domains are identical and overlap in energy [Fig. \ref{fig:fig6}(c)], whereas their intensity contributions differ due to the magnetic neutron scattering polarization factor.

Magnon splitting occurs both near the $\Gamma$ point and along the $\Gamma$--$L$ direction, but the underlying mechanisms are distinct.
Near the $\Gamma$ point in pure MnTe, the two branches are split by easy-plane single-ion anisotropy.
The gapped branch corresponds to out-of-plane spin fluctuations, which incur an energy cost due to the anisotropy.
In contrast, the gapless branch corresponds to in-plane spin fluctuations, for which rotations within the easy plane do not incur an anisotropy energy cost.
The splitting observed at intermediate wave vectors along the $\Gamma$--$L$ direction, however, originates from the altermagnetic exchange interactions \cite{liu_chiral_2024}.

\begin{figure}[!ht]
    \includegraphics[width=\columnwidth]{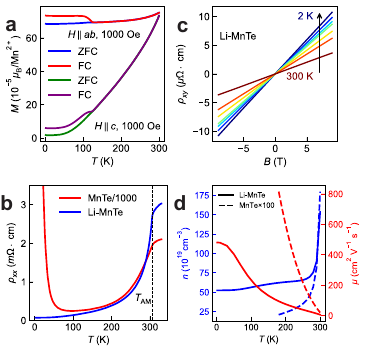}

    \caption{\label{fig:fig5}
    {\color{black}
    (a) Magnetization $M$ of Li-MnTe versus temperature $T$.
    (b) Longitudinal resistivity $\rho_{xx}$ versus $T$ for MnTe and Li-MnTe.
    (c) Transverse resistivity $\rho_{xy}$ of Li-MnTe.
    (d) Carrier density $n$ and mobility $\mu$ of MnTe and Li-MnTe.
    }}
\end{figure}

\begin{figure}[!ht]
    \includegraphics[width=\columnwidth]{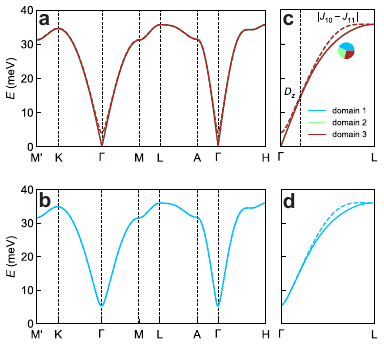}

    \caption{\label{fig:fig6}
    {\color{black}
    (a,c) Calculated spin waves of MnTe for three domains within (a) and outside (c) the nodal planes.
    The calculations use exchange parameters from Ref. \cite{liu_chiral_2024}.
    The inset in (c) displays the intensity contributions from the three domains at the position of altermagnetic splitting. The solid and dashed lines denote different branches of the spin waves, and the vertical dashed line separates the splitting dominated by single-ion anisotropy ($D_z$) and altermagnetism ($|J_{10}-J_{11}|$).
    (b,d) Calculated spin waves of Li-MnTe, with the single-ion anisotropy replaced by the value reported in Ref. \cite{yumnam_MagnonGapTuning_2024}.
    }
    }
\end{figure}

}

\clearpage
\onecolumngrid
\raggedbottom
\section*{Supplementary Information}
\renewcommand{\arraystretch}{0.6}
\renewcommand{\thetable}{S\arabic{table}}
\renewcommand{\thefigure}{S\arabic{figure}}
\renewcommand{\theHtable}{supp.\arabic{table}}
\renewcommand{\theHfigure}{supp.\arabic{figure}}
\setcounter{table}{0}
\setcounter{figure}{0}
\let\addcontentsline\suppAddContentsLine
\tableofcontents
\clearpage

\section{Experimental Details}

\subsection{Sample synthesis and characterization}
Single crystals of $\alpha$-MnTe (MnTe) and Li-doped $\alpha$-MnTe (Li-MnTe) were synthesized using Te self-flux method following the procedures as described in our previous work \cite{liu_StraintunableAnomalousHall_2025}. Magnetic susceptibility measurements were performed on a single-crystal Li-MnTe sample with a mass of 90.4 mg, using a Quantum Design Magnetic Property Measurement System (MPMS) equipped with a superconducting quantum interference device (SQUID) magnetometer. The sample was mounted on the sample holder using GE varnish. Both field-cooled (FC) and zero-field-cooled (ZFC) measurements were conducted with the magnetic field applied along the $c$-axis and within the $ab$-plane.

\subsection{Neutron diffraction}

Neutron diffraction measurements were carried out using the single crystal diffractometer TOPAZ at the Spallation Neutron Source (SNS), Oak Ridge National Laboratory (ORNL) \cite{coates2018suite}. The experiments employed a broad neutron wavelength range of 0.4--3.5~\AA{}. Single crystals of Li-MnTe ($\sim$ 100 mg) and pure MnTe ($\sim$ 50 mg) were used. The samples were mounted on the ambient goniometer, and data were collected using time-of-flight (TOF) techniques, which provided wavelength-resolved Laue diffraction patterns. Diffraction datasets for both samples were obtained at 90~K and 340~K. A nitrogen cryostream was utilized for temperature control.

The data collection strategy was determined using the planning tool in NeuXtalViz \cite{NeuXtalViz}. A multiresolution machine learning method was employed for peak integration to obtain peak intensities in three-dimensional $(H,K,L)$ space \cite{reshniak2024hierarchical}. Data reduction and normalization, including the Lorentz factor, neutron TOF spectrum, and detector efficiency corrections, were carried out following the procedure reported previously \cite{schultz2014integration}. The reduced data were saved in SHELX HKLF2 format \cite{sheldrick2015crystal} with the neutron wavelength recorded separately for each reflection. The neutron crystal structures were solved in JANA2020 \cite{petvrivcek2023jana2020} and refined successfully to convergence. Crystal structures presented in the main text were visualized using the VESTA software package \cite{vesta}.

The momentum transfer $\bm{Q}$ in 3D reciprocal space is defined as $\bm{Q} = H \bm{a}^*+K\bm{b}^*+L\bm{c}^*$ where $H$, $K$ and $L$ are Miller indices and $\bm{a}^* = 2\pi (\bm{b} \times\bm{c}) / [\bm{a} \cdot(\bm{b} \times\bm{c})]$, $\bm{b}^* = 2\pi (\bm{c} \times\bm{a}) / [\bm{a} \cdot(\bm{b} \times\bm{c})]$, $\bm{c}^* = 2\pi (\bm{a} \times\bm{b}) / [\bm{a} \cdot(\bm{b} \times\bm{c})]$ with $\bm{a} = a\hat{\bm{x}}$, $\bm{b} = a(-1/2~\hat{\bm{x}}+\sqrt{3}/2~ \hat{\bm{y}})$ and $\bm{c} = c\hat{\bm{z}}$.

\subsection{Inelastic neutron scattering}
Inelastic neutron scattering (INS) measurements on undoped MnTe and preliminary measurements on Li-MnTe were performed using the fine-resolution Fermi chopper spectrometer SEQUOIA at the Spallation Neutron Source (SNS), Oak Ridge National Laboratory (ORNL) \cite{granroth2006sequoia, granroth2010sequoia}. For the pure MnTe sample, a single crystal with a mass of approximately 0.46 g was mounted in the [$H$, 0, $L$] scattering plane. Data were collected at $T = 8$ K using an incident energy of $E_i = 50$ meV with the high-flux chopper operating at a frequency of $f = 180$ Hz. Measurements were performed over a total rotation range of 120 degrees about the vertical axis of the sample, with a step size of 0.5 degrees. During the experiment, the SNS accelerator operated at a beam power of 1.8 MW, and data were collected for an accumulated proton charge of approximately 0.4 Coulombs per angle step, corresponding to a counting time of roughly 5 minutes. The data obtained from SEQUOIA are presented in Figs. 2(b)--(e), 3(a)(c)(f)--(i), 4, and S2.

For the Li-MnTe sample, an early-stage experiment was conducted at $T = 6$ K using an array of co-aligned crystals with a total mass of approximately 0.4 g, also mounted in the [$H$, 0, $L$] scattering plane. Several experimental configurations were employed: high-flux mode with incident energy $E_i = 50$ meV at Fermi frequency $f = 300$ Hz and $E_i = 25$ meV at $f = 240$ Hz, as well as high-resolution mode with $E_i = 60$ meV at $f = 420$ Hz.

INS measurements on Li-MnTe were carried out using the TOF spectrometer MAPS at the ISIS Neutron and Muon Source, Rutherford Appleton Laboratory \cite{ewings2019MAPS}. The sample, with a mass of approximately 2.2 g, was obtained from the same large single crystal used for the TOPAZ experiment and was mounted in the [$H$, 0, $L$] scattering plane. Data were collected at $T = 5$ K with an incident energy of $E_i = 50$ meV under continuous rotation about the vertical axis of the sample. During the measurements, the accelerator operated with a synchrotron current of 152~$\mu$A on target station 1. Measurements were carried out in two modes: high-resolution (HR) mode with a Fermi chopper frequency of $f = 300$ Hz and a sample rotation range of 60 degrees, and high-flux (HF) mode with $f = 150$ Hz and a rotation range of 120~degrees. Each mode was allocated approximately 60 hours of beamtime, evenly distributed over the full angle range. The data were reduced into .nxspe files using a step of 0.5~degrees. Data from the HF mode are presented in Figs. 2(a)(c)--(e), 3(b)(d)(f)--(i), 4, and S2, while data from the HR mode are shown in Figs. 2(c)--(e), S3, and S4.

Data reduction was performed using MANTID \cite{arnold2014mantid}, Horace \cite{ewings2016horace}, and SHIVER \cite{shiver}.

\begin{figure*}[!ht]
    \includegraphics[width=0.6\linewidth]{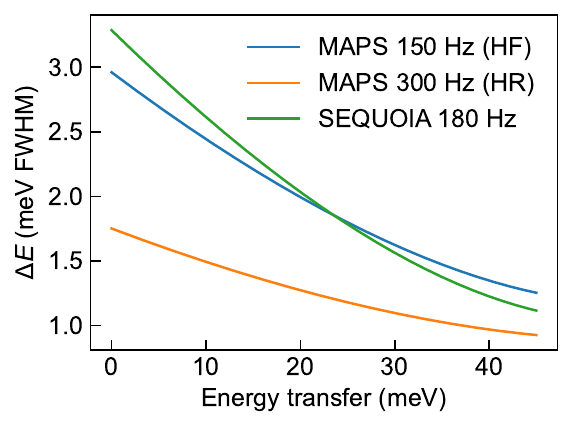}

    \caption{\label{fig:S1} Energy resolutions calculated using PyChop for the three experimental configurations used to measure the spin excitation data presented in this Letter. HF: high-flux mode; HR: high-resolution mode.
    }
\end{figure*}

\subsection{Absolute unit normalization}
INS data were normalized to absolute units during the data reduction process by measuring a standard vanadium sample of known mass \cite{10.1063/1.4818323}. The vanadium measurements were performed in the same sample environment and with the same chopper and instrument settings as the sample measurements.

\newpage

\subsection{Linear spin wave theory (LSWT) Calculations}
The spin Hamiltonian was determined from the INS spectra using linear spin wave theory (LSWT) implemented in the \textsc{SpinW} package \cite{toth2015spinw}. For both compounds, a Heisenberg spin Hamiltonian with a single-ion anisotropy term was employed, expressed as

\begin{equation}
H = \sum_{\langle i,j \rangle} J_{ij} \bm{S}_{i} \cdot \bm{S}_{j} + \sum_i D_z (S_i^z)^2,
\label{Heisenberg}
\end{equation}
where $i$ indexes the Mn$^{2+}$ ions, $\bm{S}_i$ is the spin operator at site $i$, and $S_i^z$ denotes its $z$ component. $J_{ij}$ represents the exchange interaction between spins at sites $i$ and $j$, with the summation taken over spin pairs. In this coordinate system, the $z$-axis corresponds to the crystallographic $c$-axis. The anisotropy term $D_z$ is positive for easy-plane anisotropy, as in the case of pure MnTe, and negative for Ising-type anisotropy, as observed in Li-MnTe.

To extract the exchange coupling parameters using the \textsc{SpinW} program,
we first constructed the lattice and magnetic structures and enabled the
relevant exchange interaction and single-ion anisotropy terms.
The lattice and magnetic structures were obtained from single-crystal neutron
diffraction results and CIF files available in the ICSD database.
The extracted spin-wave dispersion relation $E(\bm{Q})$ was then used
as input for the \textsc{SpinW} calculations. To determine $E(\bm{Q})$, we performed Gaussian fits to the inelastic neutron
scattering spectra at each wave vector $\bm{Q}$ along high-symmetry directions.
From these fits, we extracted the peak center energy $E$ and the half-width at
half-maximum (HWHM).

For the models ``Pure MnTe center'' and ``Li-MnTe center''
listed in Table~\ref{excp}, the fitted peak energies $E$ at each $\bm{Q}$
along the high-symmetry directions were directly used in the \textsc{SpinW}
calculations. Unless otherwise specified, all spin wave calculations presented in the main text for Li-doped MnTe are based on the exchange coupling  parameters obtained from ``Li-MnTe center'' model.

For the models ``Li-MnTe upper'' and ``Li-MnTe lower,'' we introduced a modification to account for
spin wave broadening. At the Brillouin zone (BZ) boundaries (i.e.,
along the M--L and L--A paths), we used $E+\mathrm{HWHM}$ and
$E-\mathrm{HWHM}$, respectively, to represent the upper and lower
bounds of the broadened dispersion. Along all other high-symmetry directions,
we used the unmodified fitted peak energy $E(\bm{Q})$, since the apparent broadening there is mainly
influenced by {\color{black}single-ion anisotropy}.

\begin{table}[!ht]
\centering
\caption{Exchange coupling constants calculated using \textsc{SpinW}.}
\label{excp}
\begin{tabular}{ccccccc}
\hline
Sample and data      & $J_1$     & $J_2$       & $J_3$      & $D_z$       & $J_{10}$                & $J_{11}$ (meV)           \\ \hline
Pure MnTe            & 3.8951 & -0.139   & 0.46383 & 0.05519  & -0.01378           & 0.03986             \\
Li-MnTe upper  & 3.6575 & -0.18698 & 0.47811 & -0.2411  & \multicolumn{2}{c}{\multirow{3}{*}{N/A}} \\
Li-MnTe center & 3.5909 & -0.22146 & 0.45177 & -0.24515 & \multicolumn{2}{c}{}                     \\
Li-MnTe lower  & 3.3283 & -0.21053 & 0.47816 & -0.24695 & \multicolumn{2}{c}{}                     \\ \hline
\end{tabular}
\end{table}

\begin{figure*}[!ht]
    \includegraphics[width=1.0\linewidth]{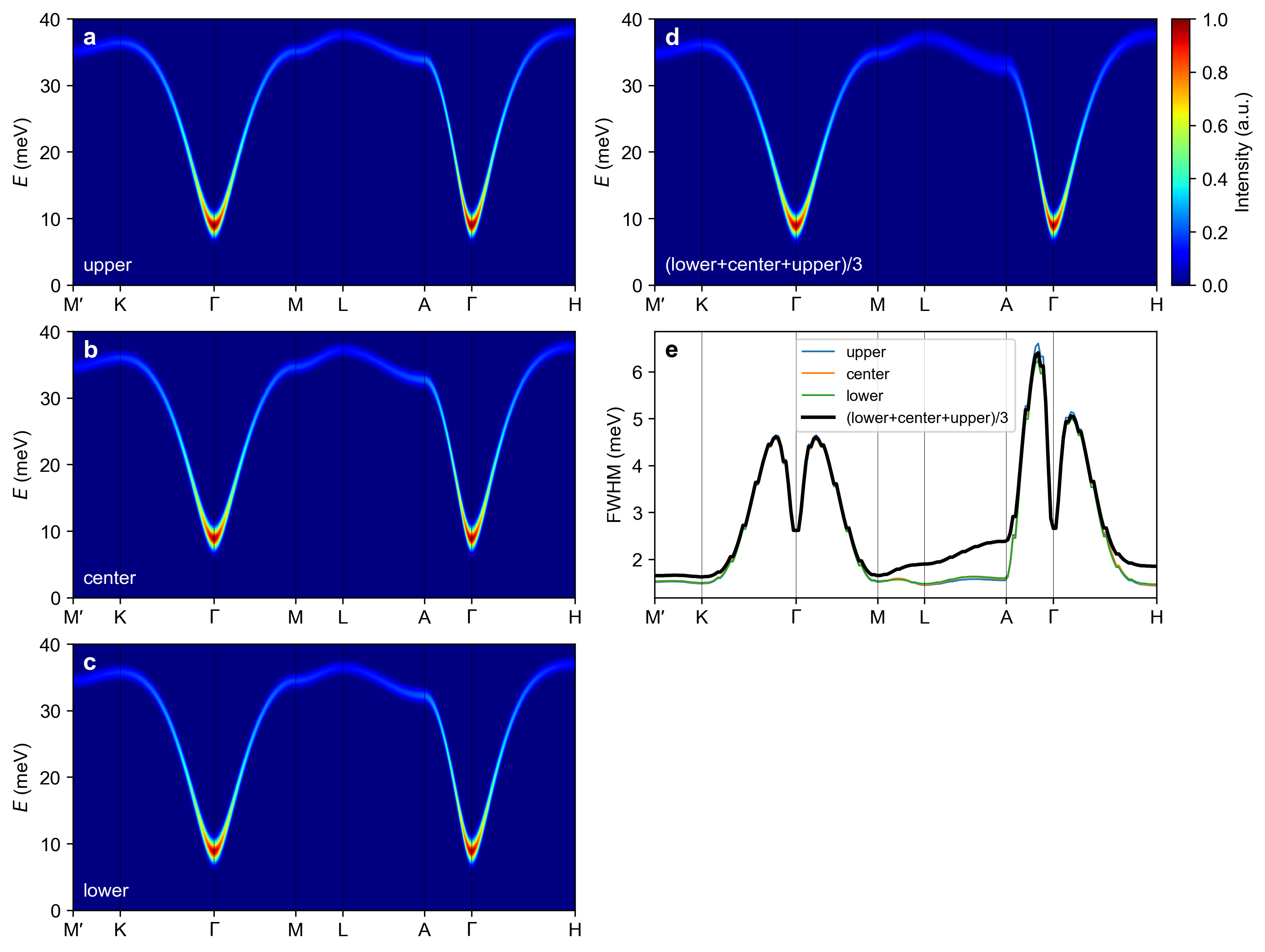}
    \caption{\label{fig:averaged_lswt_spectra} \color{black}(a-c) Calculated spin wave spectra for LSWT models ``Li-MnTe upper'', ``Li-MnTe center'', and ``Li-MnTe lower'', convolved with the resolution function in high-flux (HF) mode on MAPS. (d) Simulated spin wave spectrum resulting from a superposition of all three models with equal weights. (e) Line widths extracted from each individual model and the superimposed model, using the same integration ranges employed for the experimental data presented in Figure 2(c,d) of the main text.
    }
\end{figure*}

\newpage
\subsection{Electrical transport measurements}
The longitudinal and Hall resistivities were measured using a standard five-probe configuration. Measurements were performed in a 9~T Quantum Design PPMS DynaCool system. The voltage signals were first amplified by appropriate preamplifiers (SR554) and subsequently read out using SR830 lock-in amplifiers. The sample was polished into a bar with dimensions of about $2.0 \times 1.0 \times 0.1~\mathrm{mm}^3$. During the measurements, the magnetic field was applied along the crystallographic axis $[0,0,1]$, while the electrical current flowed along the $[1,1,0]$ direction. All field-dependent data were symmetrized and antisymmetrized to correct for contact misalignment.

\subsection{Angle-resolved photoemission spectroscopy measurements}
All ARPES data were taken at beamline 7.0.2 of the Advanced Light Source at Lawrence Berkeley National Lab using the deflector mode of a Scienta R4000 analyzer with an energy resolution of less than 20~meV. The samples were cleaved \emph{in-situ} with a base pressure better than $5 \times10^{-11}$ Torr. All measurements were performed at a temperature of 15~K using 120~eV photons with linear horizontal polarization.

\newpage
\section{Single crystal neutron diffraction}

\subsection{Experiment and refinement details}
\label{sec:diffraction}

\begin{table}[!ht]
\centering
\color{black}
\caption{\textcolor{black}{MnTe at 90 K and 340 K.
At 90 K, two magnetic structures were tested with refined occupancy.
At 340 K, refinements assuming full Te occupancy and allowing Te deficiency were compared.}}
\begin{tabular}{|c|c|c|c|c|}
\hline
$T$
& \multicolumn{2}{c|}{90 K}
& \multicolumn{2}{c|}{340 K} \\
\hline

Space group
& \multicolumn{4}{c|}{$P6_3/mmc$ (No. 194)} \\
\hline

Magnetic space group
& $Cm'c'm$
& $Cmcm$
& \multicolumn{2}{c|}{N/A} \\
\hline

Te occupancy treatment
& \multicolumn{2}{c|}{---}
& Te deficient
& Te full occ. \\
\hline

$a$ (\AA)
& \multicolumn{2}{c|}{4.1300(5)}
& \multicolumn{2}{c|}{4.1522(5)} \\
\hline

$c$ (\AA)
& \multicolumn{2}{c|}{6.6530(11)}
& \multicolumn{2}{c|}{6.7162(11)} \\
\hline

$V$ (\AA$^3$)
& \multicolumn{2}{c|}{98.28(2)}
& \multicolumn{2}{c|}{100.28(2)} \\
\hline

Index range
& \multicolumn{2}{c|}{
    \begin{tabular}{@{}c@{}}
    $-8 < H < 9$ \\
    $-8 < K < 9$ \\
    $-14 < L < 14$
    \end{tabular}
}
& \multicolumn{2}{c|}{
    \begin{tabular}{@{}c@{}}
    $-9 < H < 8$ \\
    $-8 < K < 9$ \\
    $-14 < L < 14$
    \end{tabular}
} \\
\hline

\begin{tabular}{@{}c@{}}
Reflections collected \\
($N_{\mathrm{obs}}/N_{\mathrm{all}}$)
\end{tabular}
& \multicolumn{2}{c|}{2137/2154}
& \multicolumn{2}{c|}{1460/1468} \\
\hline

\begin{tabular}{@{}c@{}}
$R$ indices \\
($I > 3\sigma(I)$)
\end{tabular}
& \begin{tabular}{@{}c@{}}
  $R = 3.57$ \\
  $wR2 = 10.89$
  \end{tabular}
& \begin{tabular}{@{}c@{}}
  $R = 3.47$ \\
  $wR2 = 9.97$
  \end{tabular}
& \begin{tabular}{@{}c@{}}
  $R = 3.13$ \\
  $wR2 = 7.69$
  \end{tabular}
& \begin{tabular}{@{}c@{}}
  $R = 3.19$ \\
  $wR2 = 8.25$
  \end{tabular} \\
\hline

$R$ indices (all data)
& \begin{tabular}{@{}c@{}}
  $R = 3.63$ \\
  $wR2 = 10.92$
  \end{tabular}
& \begin{tabular}{@{}c@{}}
  $R = 3.52$ \\
  $wR2 = 10.01$
  \end{tabular}
& \begin{tabular}{@{}c@{}}
  $R = 3.16$ \\
  $wR2 = 7.70$
  \end{tabular}
& \begin{tabular}{@{}c@{}}
  $R = 3.21$ \\
  $wR2 = 8.25$
  \end{tabular} \\
\hline

Goodness-of-fit
& 2.84
& 2.16
& 1.85
& 1.99 \\
\hline

Moment size ($\mu_{\mathrm{B}}$)
& 4.35(3)
& 4.30(3)
& \multicolumn{2}{c|}{N/A} \\
\hline
\end{tabular}
\end{table}

For both MnTe and Li-MnTe, we find no evidence of any other impurity phases, clearly different from powder samples of Li-MnTe \cite{yumnam_MagnonGapTuning_2024}.
For MnTe, we also evaluated the possibility of intersite mixing, specifically the occupation of Te sites by Mn atoms, using the neutron diffraction dataset collected at 340 K. However, this model did not yield an improved refinement compared to the current model, which assumes that defects are restricted to Te sites and that Mn atoms exclusively occupy their designated lattice positions. {\color{black} We note that Te vacancies are thermodynamically unfavorable during crystal growth in a Te flux, and thus the observed Te deficiency is likely an artifact of the data quality.}

To determine the magnetic structure, two models were tested using the dataset obtained at 90 K. The first model adopts the magnetic space group $Cmcm$, where the magnetic moments align along the crystallographic $a$ axis. The second model corresponds to the magnetic space group $Cm'c'm$, in which the moments are oriented along the reciprocal $a^*$ axis. Although the $Cmcm$ model yielded a slightly better refinement result, the $Cm'c'm$ model is more consistent with our previous neutron diffraction results under strain-induced detwinning of magnetic domains \cite{liu_StraintunableAnomalousHall_2025}.

\begin{table}[!t]
\centering
\color{black}
\small
\setlength{\tabcolsep}{3.8pt}
\caption{
Li-MnTe at 340 K. Models placing Li at Wyckoff site $2a$, $2c$, $2d$, $4f$, or uncorrelated interstitial sites were tested.
}
\label{LiMnTe340K_refine}
\vspace{0.5em}

\scalebox{0.675}{
\begin{tabular}{|c|c|c|c|c|c|}
\hline
$T$
& \multicolumn{5}{c|}{340 K} \\
\hline
Space group
& \multicolumn{5}{c|}{$P6_3/mmc$ (No. 194)} \\
\hline
Li site model
& \begin{tabular}{@{}c@{}} Uncorrelated \end{tabular}
& \begin{tabular}{@{}c@{}} Mn site \end{tabular}
& \begin{tabular}{@{}c@{}} Te site \end{tabular}
& \begin{tabular}{@{}c@{}} Interstitial $2d$ \end{tabular}
& \begin{tabular}{@{}c@{}} Interstitial $4f$ \end{tabular} \\
\hline
Li Wyckoff position
& ---
& $2a$ $(0, 0, 0)$
& $2c$ $(1/3, 2/3, 1/4)$
& $2d$ $(1/3, 2/3, 3/4)$
& $4f$ $(1/3, 2/3, 0.902(16))$ \\
\hline
Refined Li occupancy
& None
& 0.030(7)
& 0.029(3)
& 0.001(14)
& 0.013(13) \\
\hline
Refined Te occupancy
& 0.961(3)
& 0.947(3)
& 0.971(3)
& 0.961(5)
& 0.960(4) \\
\hline
Chemical formula
& $\mathrm{MnTe}_{0.961(3)}$
& $(\mathrm{Mn}_{0.970(7)}\mathrm{Li}_{0.030(7)})\mathrm{Te}_{0.947(3)}$
& $\mathrm{Mn}(\mathrm{Te}_{0.971(3)}\mathrm{Li}_{0.029(3)})$
& $\mathrm{Li}_{0.001(14)}\mathrm{MnTe}_{0.961(5)}$
& $\mathrm{Li}_{0.013(13)}\mathrm{MnTe}_{0.960(4)}$ \\
\hline
$a$ (\AA)
& \multicolumn{5}{c|}{4.1532(3)} \\
\hline
$c$ (\AA)
& \multicolumn{5}{c|}{6.7190(8)} \\
\hline
$V$ (\AA$^3$)
& \multicolumn{5}{c|}{100.37(2)} \\
\hline
Index range
& \multicolumn{5}{c|}{
    \begin{tabular}{@{}c@{}}
    $-8 \le H \le 8$ \\
    $-6 \le K \le 7$ \\
    $-12 \le L \le 13$
    \end{tabular}
} \\
\hline
\begin{tabular}{@{}c@{}}
Reflections collected \\
($N_{\mathrm{obs}}/N_{\mathrm{all}}$)
\end{tabular}
& \multicolumn{5}{c|}{1991 / 2015} \\
\hline
Extinction coeff. $G_{\mathrm{iso}}$
& 0.00115(4)
& 0.00129(5)
& 0.00125(5)
& 0.00125(5)
& 0.00125(5) \\
\hline
\begin{tabular}{@{}c@{}}
$R$ indices \\
($I > 3\sigma(I)$)
\end{tabular}
& \begin{tabular}{@{}c@{}}
  $R = 4.59\%$ \\
  $wR_2 = 15.31\%$
  \end{tabular}
& \begin{tabular}{@{}c@{}}
  $R = 4.59\%$ \\
  $wR_2 = 15.31\%$
  \end{tabular}
& \begin{tabular}{@{}c@{}}
  $R = 4.59\%$ \\
  $wR_2 = 15.32\%$
  \end{tabular}
& \begin{tabular}{@{}c@{}}
  $R = 4.59\%$ \\
  $wR_2 = 15.31\%$
  \end{tabular}
& \begin{tabular}{@{}c@{}}
  $R = 4.59\%$ \\
  $wR_2 = 15.30\%$
  \end{tabular} \\
\hline
\begin{tabular}{@{}c@{}}
$R$ indices \\
(all data)
\end{tabular}
& \begin{tabular}{@{}c@{}}
  $R = 4.66\%$ \\
  $wR_2 = 15.33\%$
  \end{tabular}
& \begin{tabular}{@{}c@{}}
  $R = 4.66\%$ \\
  $wR_2 = 15.33\%$
  \end{tabular}
& \begin{tabular}{@{}c@{}}
  $R = 4.66\%$ \\
  $wR_2 = 15.33\%$
  \end{tabular}
& \begin{tabular}{@{}c@{}}
  $R = 4.65\%$ \\
  $wR_2 = 15.33\%$
  \end{tabular}
& \begin{tabular}{@{}c@{}}
  $R = 4.66\%$ \\
  $wR_2 = 15.31\%$
  \end{tabular} \\
\hline
Goodness-of-fit ($\mathrm{GOF}$)
& 3.90
& 3.90
& 3.90
& 3.90
& 3.90 \\
\hline
Mn $U_{11} = U_{22}$ (\AA$^2$)
& 0.0187(2)
& 0.0187(2)
& 0.0188(2)
& 0.0187(2)
& 0.0187(2) \\
\hline
Mn $U_{33}$ (\AA$^2$)
& 0.0155(3)
& 0.0155(3)
& 0.0155(3)
& 0.0155(3)
& 0.0156(3) \\
\hline
Te $U_{11} = U_{22}$ (\AA$^2$)
& 0.0091(1)
& 0.0091(1)
& 0.0091(1)
& 0.0091(1)
& 0.0091(1) \\
\hline
Te $U_{33}$ (\AA$^2$)
& 0.0119(2)
& 0.0119(2)
& 0.0119(2)
& 0.0119(2)
& 0.0118(2) \\
\hline
Li $U_{11} = U_{22}$ (\AA$^2$)
& ---
& 0.0187(2)$^*$
& 0.0091(1)$^\dagger$
& \multicolumn{2}{c|}{---} \\
\hline
Li $U_{33}$ (\AA$^2$)
& ---
& 0.0155(3)$^*$
& 0.0119(2)$^\dagger$
& \multicolumn{2}{c|}{---} \\
\hline
Li $U_{\mathrm{iso}}$ (\AA$^2$)
& \multicolumn{3}{c|}{---}
& 0.017(25)
& 0.011(27) \\
\hline
\multicolumn{6}{l}{\footnotesize $^*$Tied to Mn1 ($u_{11}, u_{33}$). \quad $^\dagger$Tied to Te1 ($u_{11}, u_{33}$).}
\end{tabular}
}
\end{table}

We used the 340 K diffraction dataset to investigate the possible crystallographic location of Li atoms.
As a first scenario, we considered the possibility that Li atoms occupy uncorrelated interstitial sites. In this case, the Li atoms would not contribute coherent scattering, and therefore the Bragg peak intensities would remain unaffected. Accordingly, Li atoms were not included in the refinement model for this scenario.

{\color{black}
We then tested alternative structural models incorporating Li atoms at different crystallographic sites. Stable refinements were obtained when Li was placed at the $4f$, $2d$, and $2c$ positions (the latter coinciding
with the Te site). However, when Li atoms were assigned to the $2d$ site, the refined concentration was significantly lower than the nominal doping level of 5\%, as shown in Table~\ref{Li2d340K}. A similar refinement using the 90~K dataset yielded negative Li occupancies at the $2d$ site when the atomic displacement parameters (ADPs) were not constrained, which is physically implausible. Placing Li at the $4f$ site resulted in a refined concentration of $\sim$1.3\%, still much smaller than the nominal doping level. In contrast, assigning Li atoms to the 2$c$ site resulted in stable refinements at both temperatures, with a consistent refined concentration of approximately 3\%, though this site is chemically implausible. We also attempted to refine the model with Li on the Mn $2a$ site, with a fixed, representative Li occupancy of 3\%, which yielded identical fit quality with a slightly higher Te deficit. Nevertheless, we note that the choice of site does not significantly change the overall refinement quality (Table~\ref{LiMnTe340K_refine}). Therefore, we cannot conclusively establish the location of Li within the structure using the current data.
}

\begin{table}[!ht]
\centering
\caption{Li-MnTe at 90 K.}
\begin{tabular}{|c|clc|}
\hline
Magnetic space group                                                                          & \multicolumn{2}{c|}{$P6_3'/m'm'c$}                                                         & $Cm'cm'$                                                                               \\ \hline
Space group                                                                      & \multicolumn{3}{c|}{$P6_3/mmc$ (No. 194)}                                                                                                                             \\ \hline
$a$                                                                                & \multicolumn{3}{c|}{4.130(2)}                                                                                                                                     \\ \hline
$c$                                                                                & \multicolumn{3}{c|}{6.666(2)}                                                                                                                                     \\ \hline
$V$                                                                                & \multicolumn{3}{c|}{98.48(7)}                                                                                                                                   \\ \hline
Index range                                                                      & \multicolumn{3}{c|}{\begin{tabular}[c]{@{}c@{}}-9\textless{}$H$\textless{}9\\ -7\textless{}$K$\textless{}8\\ -13\textless{}$L$\textless{}14\end{tabular}}                \\ \hline
\begin{tabular}[c]{@{}c@{}}Reflections collected\\ ($N$(obs)/$N$(all))\end{tabular}  & \multicolumn{3}{c|}{3215/3261}                                                                                                                                     \\ \hline
\begin{tabular}[c]{@{}c@{}}$R$ indices \\ ($I$ \textgreater 3$\sigma(I)$)\end{tabular} & \multicolumn{2}{c|}{\begin{tabular}[c]{@{}c@{}}$R$=4.65\\ $wR2$=16.04\end{tabular}} & \begin{tabular}[c]{@{}c@{}}$R$=4.71\\ $wR2$ = 16.14\end{tabular}                     \\ \hline
$R$ indices (all data)                                                             & \multicolumn{2}{c|}{\begin{tabular}[c]{@{}c@{}}$R$=4.71\\ $wR2$=16.06\end{tabular}}   & \begin{tabular}[c]{@{}c@{}}$R$ = 4.76\\ $wR2$=16.16\end{tabular}                     \\ \hline
Goodness-of-fit                                                                  & \multicolumn{2}{c|}{4.27}                                                       & 4.30                                                                             \\ \hline

Moment size ($\mu_\mathrm{B}$)                                                                  & \multicolumn{2}{c|}{4.29(3)}                                                       & 4.30(10)                                                                             \\ \hline

\end{tabular}
\label{LiMnTe90}
\end{table}

In accordance with previous powder neutron diffraction studies \cite{moseley_giant_2022}, we applied the orthorhombic magnetic space group $Cm'cm'$ to refine the magnetic structure of Li-MnTe. This model also yielded a vanishingly small in-plane magnetic moment at the given doping level. In addition, we tested a hexagonal magnetic space group with higher symmetry, in which the magnetic structure is strictly antiferromagnetic and the magnetic moments are constrained to lie along the crystallographic $c$ axis. The refinement results for both models are summarized in Table~\ref{LiMnTe90}.

\newpage
\subsection{Fractional atomic coordinates and displacement parameter (\AA$^2$) determined from neutron diffraction}

\begin{table}[!ht]
\centering
\caption{MnTe, 340 K}
\begin{tabular}{ccccccccc}
\hline
Atom & Label & $x$   & $y$   & $z$   & Occ.  & $U_{\mathrm{iso}}/U_{\mathrm{eq}}$     & Site & Sym. \\ \hline
Mn   & Mn1   & 0   & 0   & 0   & 1.000 & 0.017 & 2$a$   & $\bar{3}m$  \\
Te   & Te1   & 1/3 & 2/3 & 1/4 & 0.967(2) & 0.010 & 2$c$   & $\bar{6}m2$ \\ \hline
\end{tabular}
\end{table}

\begin{table}[!ht]
\centering
\caption{MnTe, 90 K}
\begin{tabular}{ccccccccc}
\hline
Atom & Label & $x$   & $y$   & $z$   & Occ.  & $U_{\mathrm{iso}}/U_{\mathrm{eq}}$     & Site & Sym. \\ \hline
Mn   & Mn1   & 0   & 0   & 0   & 1.000 & 0.004 & 2$a$   & $\bar{3}m$  \\
Te   & Te1   & 1/3 & 2/3 & 1/4 & 0.957(2) & 0.002 & 2$c$   & $\bar{6}m2$ \\ \hline
\end{tabular}
\end{table}

\begin{table}[!ht]
\centering
\caption{Li-MnTe, 340 K without Li atoms}
\begin{tabular}{ccccccccc}
\hline
Atom & Label & $x$   & $y$   & $z$   & Occ.  & $U_{\mathrm{iso}}/U_{\mathrm{eq}}$     & Site & Sym. \\ \hline
Mn   & Mn1   & 0   & 0   & 0   & 1.000 & 0.017 & 2$a$   & $\bar{3}m$  \\
Te   & Te1   & 1/3 & 2/3 & 1/4 & 0.961(3) & 0.010 & 2$c$   & $\bar{6}m2$ \\ \hline
\end{tabular}
\end{table}

\begin{table}[!ht]
\centering
\caption{Li-MnTe, 340 K with the assumption at Li atoms sit at the Te site}
\begin{tabular}{ccccccccc}
\hline
Atom & Label & $x$   & $y$   & $z$   & Occ.  & $U_{\mathrm{iso}}/U_{\mathrm{eq}}$     & Site & Sym. \\ \hline
Mn   & Mn1   & 0   & 0   & 0   & 1.000 & 0.017 & 2$a$   & $\bar{3}m$  \\
Te   & Te1   & 1/3 & 2/3 & 1/4 & 0.970(1)  & 0.010 & 2$c$   & $\bar{6}m2$ \\
Li   & Li1   & 1/3 & 2/3 & 1/4 & 0.030(1)  & 0.010 & 2$c$   & $\bar{6}m2$ \\ \hline
\end{tabular}
\end{table}

\begin{table}[!ht]
\centering
\caption{Li-MnTe, 340 K with the assumption at Li atoms sit at the 2$d$ site}
\begin{tabular}{ccccccccc}
\hline
Atom & Label & $x$   & $y$   & $z$   & Occ.  & $U_{\mathrm{iso}}/U_{\mathrm{eq}}$     & Site & Sym. \\ \hline
Mn   & Mn1   & 0   & 0   & 0   & 1.000 & 0.018 & 2$a$   & $\bar{3}m$ \\
Te   & Te1   & 1/3 & 2/3 & 1/4 & 0.962(4)  & 0.010 & 2$c$   & $\bar{6}m2$ \\
Li   & Li1   & 1/3 & 2/3 & 3/4 & 0.007(16)  & 0.041 & 2$d$   & $\bar{6}m2$ \\ \hline
\end{tabular}
\label{Li2d340K}
\end{table}

\begin{table}[!ht]
\centering
\caption{Li-MnTe, 340 K with the assumption at Li atoms sit at the 4$f$ site}
\begin{tabular}{ccccccccc}
\hline
Atom & Label & $x$   & $y$   & $z$   & Occ.  & $U_{\mathrm{iso}}/U_{\mathrm{eq}}$     & Site & Sym. \\ \hline
Mn   & Mn1   & 0   & 0   & 0   & 1.000 & 0.018 & 2$a$   & $\bar{3}m$ \\
Te   & Te1   & 1/3 & 2/3 & 1/4 & 0.959(3)  & 0.010 & 2$c$   & $\bar{6}m2$ \\
Li   & Li1   & 1/3 & 2/3 & 0.898(14) & 0.009(6)  & 0.02 & 4$f$   & $3m$ \\ \hline
\end{tabular}
\label{Li4f340K}
\end{table}

\begin{table}[!ht]
\centering
\caption{Li-MnTe, 90 K with the assumption at Li atoms sit at the Te site}
\begin{tabular}{ccccccccc}
\hline
Atom & Label & $x$   & $y$   & $z$   & Occ.  & $U_{\mathrm{iso}}/U_{\mathrm{eq}}$     & Site & Sym. \\ \hline
Mn   & Mn1   & 0   & 0   & 0   & 1.000 & 0.004 & 2$a$   & $\bar{3}m'$  \\
Te   & Te1   & 1/3 & 2/3 & 1/4 & 0.969(2)  & 0.002 & 2$c$   & $\bar{6}'m'2$ \\
Li   & Li1   & 1/3 & 2/3 & 1/4 & 0.031(2)  & 0.002 & 2$c$   & $\bar{6}'m'2$ \\ \hline
\end{tabular}
\end{table}

\newpage
\section{INS data for Li-MnTe taken using the HR mode}

\begin{figure*}[!ht]
    \includegraphics[width=0.6\linewidth]{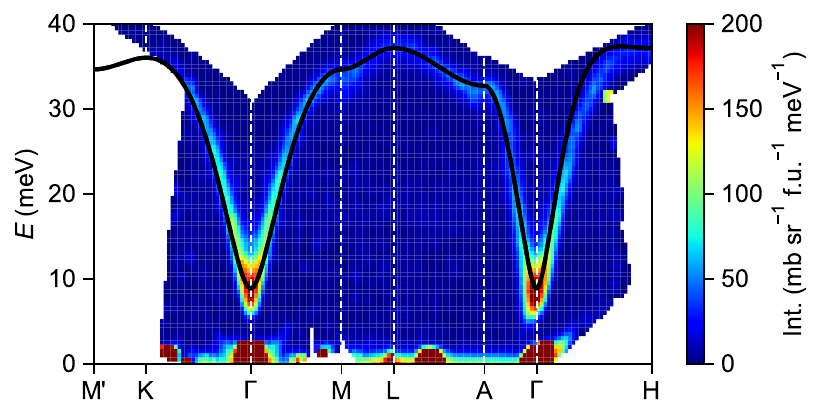}

    \caption{\label{fig:S2} Spin waves in Li-MnTe along high-symmetry directions in high-resolution mode.
    }
\end{figure*}

\begin{figure*}[!ht]
    \includegraphics[width=\linewidth]{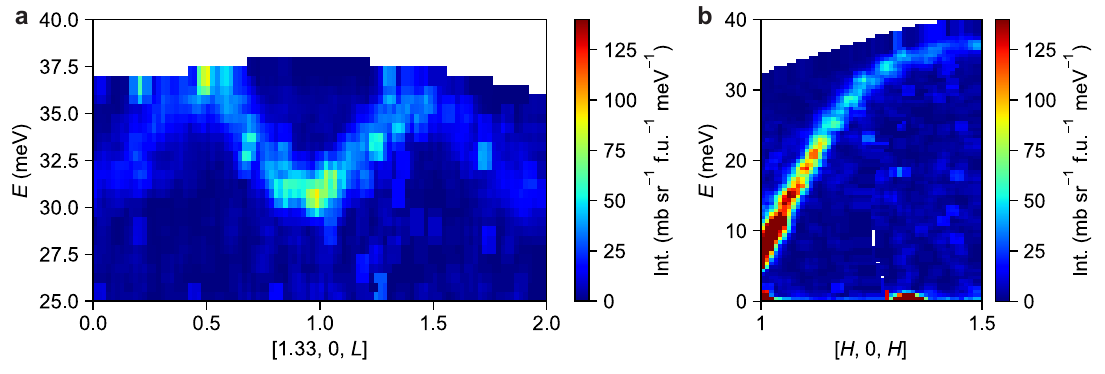}

    \caption{\label{fig:S3} (a) INS spectra of Li-MnTe along the $[1.33,0,L]$ and (b) $[H,0,H]$ directions in high-resolution mode.
    }
\end{figure*}

\newpage
\section{Energy-dependence of magnon lifetime in Li-MnTe}

\begin{figure*}[!ht]
    \includegraphics[width=0.9\linewidth]{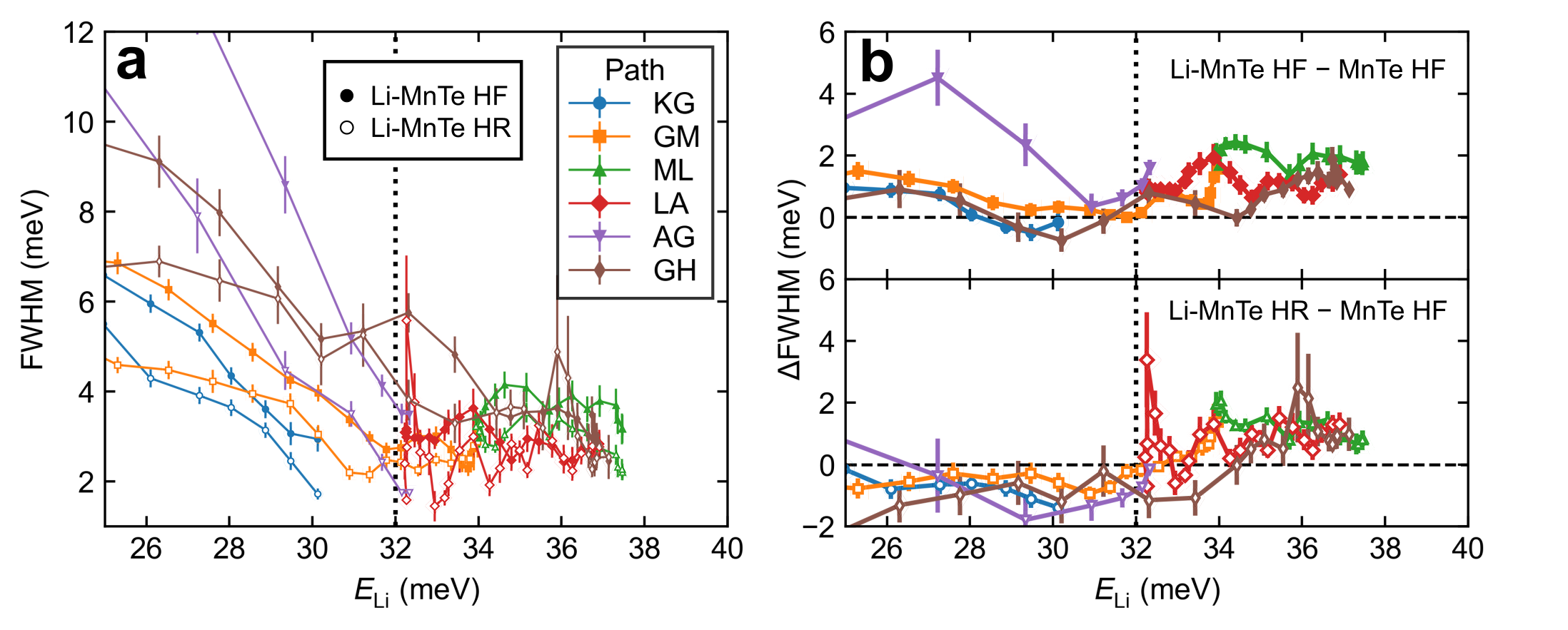}

    \caption{\label{fig:fwhm_edep} (a) Energy-dependence of the full width at half-maximum (FWHM) for Li-MnTe in high-flux (HF) mode (filled symbols) and high-resolution (HR) mode (open symbols). Dashed line serves as a guide to the eye marking the onset of increased linewidth above $\approx32$~meV. (b) Differences in FWHM ($\Delta$FWHM) between Li-MnTe and MnTe along various high-symmetry directions in the nodal planes of the BZ. The top panel utilizes HF data for both Li-MnTe and MnTe, where the resolution is comparable between Li-MnTe and MnTe measurements. The bottom panel uses HR data for Li-MnTe and HF data for MnTe, meaning the resolutions are different for each.
    FWHM values were subtracted at identical $\bm{Q}$ for both samples, and the corresponding energy of the Li-MnTe dataset ($E_\mathrm{Li}$) is used as the $x$-axis value for Beach point.}
\end{figure*}

\newpage
\section{Additional transport data}

\begin{figure*}[!ht]
    \includegraphics[width=0.9\linewidth]{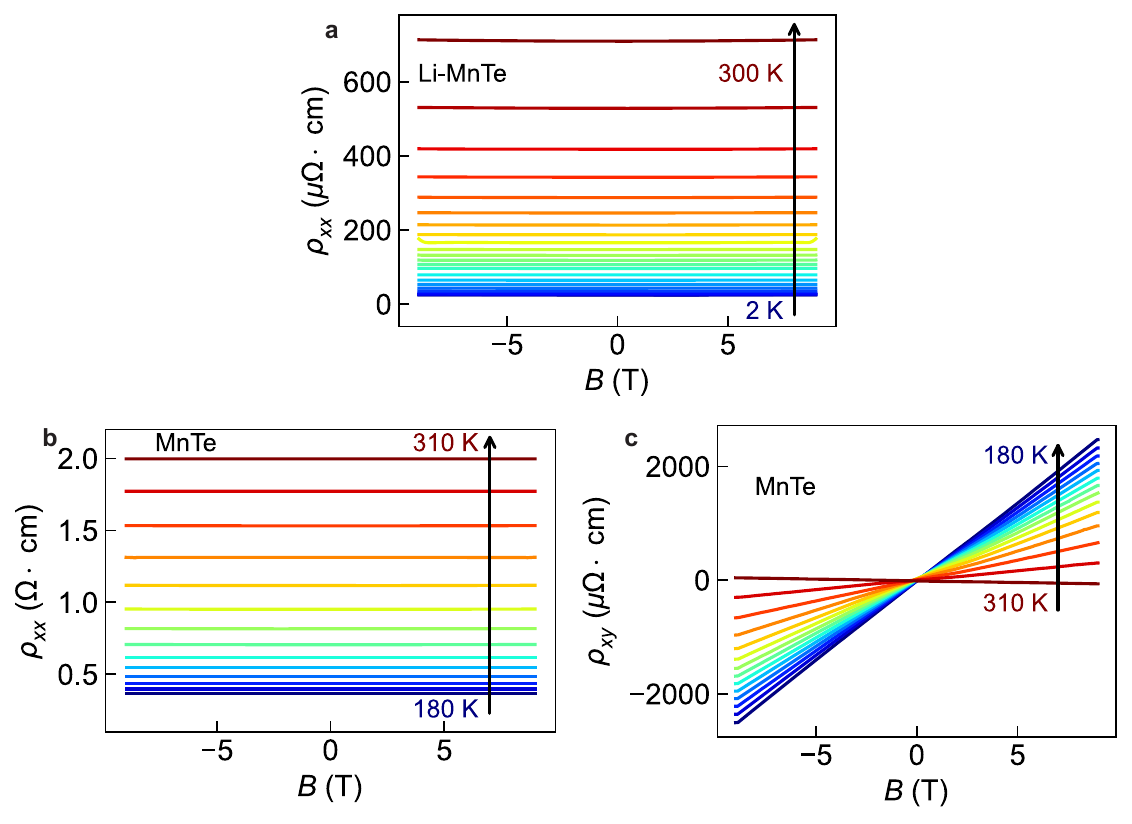}

    \caption{\label{fig:S4} Longitudinal and transverse resistivities $\rho_{xx}$ and $\rho_{xy}$ of pure and Li-MnTe.
    }
\end{figure*}

\newpage
{\color{black}
\section{Local moment sum rule}

\subsection{Theory}
The total moment sum rule asserts that the total magnetic moment $M_\mathrm{tot}$ is related to its ordered component $M_\mathrm{ord}$ and its dynamic component $M_\mathrm{dyn}$ via
\begin{equation}
    M_\mathrm{tot}^2 = M_\mathrm{ord}^2 + M_\mathrm{dyn}^2.
    \label{eqn:sumrule_def}
\end{equation}
\noindent
The ordered moment may be obtained from neutron diffraction measurements, as detailed in Sec.~\ref{sec:diffraction}. The dynamic moment may be obtained from inelastic neutron scattering measurements, as described in Ref.~\cite{10.1063/1.4818323} for the general case, and reproduced here with additional considerations for the present analysis of MnTe and Li-MnTe.

We begin our discussion with the dynamic spin correlation function
\begin{equation}
    S^{\alpha\beta}(\bm{Q},\omega) = \frac{1}{2\pi}\int_{-\infty}^{\infty} \mathrm{d}t\, e^{-i \omega t} \sum_l e^{i \bm{Q} \cdot \mathbf{r}_l} \langle S_0^\alpha(0) S_l^\beta(t) \rangle,
    \label{eqn:spin_corr}
\end{equation}
\noindent
where $\alpha,\beta$ are the Cartesian directions (e.g., $x,y,z$), $\bm{Q}$ is the momentum transfer, $\hbar\omega = E$ is the energy transfer, and $\mathbf{r}_l$ is the position of the magnetic site $l$ relative to site $0$.
Isolating the $l = 0$ on-site correlation $\langle S_0^\alpha(0) S_0^\beta(0) \rangle$ (i.e., the local moment) via Fourier
orthogonality requires integrating over the single-ion Mn$^{2+}$ sublattice Brillouin zone ($\mathrm{BZ}_\mathrm{sub}$) for which
\begin{equation}
    \frac{\int_{\mathrm{BZ}_\mathrm{sub}} e^{i\bm{Q}\cdot\mathbf{r}_l}\,\mathrm{d}\bm{Q}}{\int_\mathrm{BZ_{sub}}\mathrm{d}\bm{Q}} =
    \delta_{l,0}
\end{equation}
vanishes for all non-zero Mn--Mn separations $\mathbf{r}_l \neq \mathbf{0}$, yielding the
single-ion sum rule
\begin{equation}
    \frac{\int^\infty_{-\infty}  \int_{\mathrm{BZ}_\mathrm{sub}}
    S^{\alpha\beta}(\bm{Q},\omega)\, \mathrm{d}\bm{Q} \mathrm{d}\omega}{\int_{\mathrm{BZ}_\mathrm{sub}} \mathrm{d}\bm{Q}} =
    \frac{1}{3}S(S+1)\delta_{\alpha\beta},
    \label{eqn:spin_sumrule}
    \end{equation}
    \noindent
    where $S$ is the spin quantum number ($S=5/2$ for Mn$^{2+}$) and $\delta_{\alpha\beta}$ is the Kronecker
    delta.

For the present time-of-flight inelastic neutron scattering measurements on SEQUOIA (MnTe) and MAPS (Li-MnTe), the momentum- and energy-dependent scattered intensity $I(\bm{Q},E)$ after data reduction and normalization by vanadium standard incoherent elastic scattering is
\begin{equation}
    I(\bm{Q},E) = \frac{k_i}{k_f} \frac{1}{N} \frac{\mathrm{d}^2\sigma}{\mathrm{d}\Omega \mathrm{d}E}
    \label{eqn:reduced_intensity}
\end{equation}
\noindent
in absolute units of b~sr$^{-1}$~f.u.$^{-1}$~meV$^{-1}$, where $k_i$ and $k_f$ are the incident and final neutron wavevectors, $N$ is the total number of formula units (f.u.) in the beam, and $\frac{d^2\sigma}{d\Omega dE}$ is the partial differential magnetic scattering cross section. The latter is related to the dynamic spin correlation function by
\begin{equation}
    \frac{\mathrm{d}^2\sigma}{\mathrm{d}\Omega \mathrm{d}E} = \frac{N}{\hbar} \frac{k_f}{k_i} \left(\frac{g \gamma r_0}{2}\right)^2 f(|\bm{Q}|)^2 e^{-2W} \sum_{\alpha,\beta} (\delta_{\alpha\beta}-\tilde{Q}_\alpha\tilde{Q}_\beta)S^{\alpha\beta}(\bm{Q},\omega),
    \label{eqn:crosssection}
\end{equation}
\noindent
where $g \approx 2$ is the Land\'e $g$-factor, $\gamma = -1.913$ is the neutron gyromagnetic ratio in nuclear magnetons, $r_0 = 2.818 \times 10^{-15}$~m is the classical electron radius [giving $(\gamma r_0 / 2)^2 = 0.07265$~b], $f(|\bm{Q}|)$ is the Mn$^{2+}$ magnetic form factor, $e^{-2W}$ is the Debye-Waller factor (taken as $e^{-2W} \approx 1$ at low temperature), and $\tilde{Q}_\alpha = Q_\alpha / |\bm{Q}|$ are the projections of the unit wavevector $\tilde{\bm{Q}}$ (parallel to $\bm{Q}$) onto the Cartesian axes.
The polarization factor $(\delta_{\alpha\beta}-\tilde{Q}_\alpha\tilde{Q}_\beta)$ accounts for the fact that neutrons are sensitive only to the magnetic moment components (whether static or dynamic) perpendicular to the momentum transfer $\bm{Q}$ during a scattering event.

To match the energy units of the reduced experimental data, we define the energy-normalized dynamic spin correlation function $S^{\alpha\beta}(\bm{Q},E) \equiv \hbar^{-1} S^{\alpha\beta}(\bm{Q},\omega)$ in units of meV$^{-1}$, along with the polarization-contracted dynamic structure factor
\begin{equation}
    \tilde{S}(\bm{Q},E) \equiv \sum_{\alpha,\beta} (\delta_{\alpha\beta}-\tilde{Q}_\alpha\tilde{Q}_\beta)S^{\alpha\beta}(\bm{Q},E) = \frac{I(\bm{Q},E)}{\left(\frac{g \gamma r_0}{2}\right)^2 f(|\bm{Q}|)^2 e^{-2W}}.
    \label{eqn:stilde}
\end{equation}
\noindent
In the magnetically ordered state, inelastic spin-wave excitations correspond to transverse spin fluctuations perpendicular to the local ordered moment direction, whereas longitudinal components parallel to the ordered moment reside in the elastic Bragg channel. Assuming isotropic transverse fluctuations between the two orthogonal transverse axes ($\perp_1$ and $\perp_2$) such that $S^{\perp_1\perp_1}(\bm{Q},E) = S^{\perp_2\perp_2}(\bm{Q},E) \equiv S_\perp(\bm{Q},E)$ and vanishing off-diagonal elements, Eq.~(\ref{eqn:stilde}) simplifies to
\begin{equation}
    \tilde{S}(\bm{Q},E) = P(\bm{Q}) S_\perp(\bm{Q},E),
    \label{eqn:stilde_pol}
\end{equation}
\noindent
where $P(\bm{Q}) = (1 - \tilde{Q}_{\perp_1}^2) + (1 - \tilde{Q}_{\perp_2}^2)$ is the effective polarization factor summed over both transverse fluctuation modes.

For pristine MnTe, the ordered moments lie in the basal $ab$-plane along the crystallographic next-nearest-neighbor Mn--Mn bonds, forming three symmetry-equivalent magnetic domains separated by $120^\circ$. Within each domain, one transverse fluctuation mode lies in the $ab$-plane ($\perp,ab$) and the other lies along the $c$-axis ($z$). Averaging equally over the three $120^\circ$ domains yields $\langle 1 - \tilde{Q}_{\perp,ab}^2 \rangle_\mathrm{dom} = \frac{1}{2}(1 + \tilde{Q}_z^2)$ for the in-plane transverse mode and $1 - \tilde{Q}_z^2$ for the $c$-axis mode, where $\tilde{Q}_z = Q_c / |\bm{Q}|$. The resulting domain-averaged polarization factor is
\begin{equation}
    P_\mathrm{MnTe}(\bm{Q}) = \frac{1}{2}\left(1 + \tilde{Q}_z^2\right) + \left(1 - \tilde{Q}_z^2\right) = \frac{3}{2} - \frac{1}{2}\tilde{Q}_z^2.
    \label{eqn:pol_pure}
\end{equation}
\noindent
In contrast, for Li-MnTe, the ordered moments point along the crystallographic $c$-axis ($z$), so both transverse fluctuation modes ($x$ and $y$) lie within the $ab$-plane. Using $\tilde{Q}_x^2 + \tilde{Q}_y^2 + \tilde{Q}_z^2 = 1$, the polarization factor requires no domain averaging and is given by
\begin{equation}
    P_{\mathrm{Li\text{-}MnTe}}(\bm{Q}) = \left(1 - \tilde{Q}_x^2\right) + \left(1 - \tilde{Q}_y^2\right) = 1 + \tilde{Q}_z^2.
    \label{eqn:pol_li}
\end{equation}

Dividing $\tilde{S}(\bm{Q},E)$ by the appropriate polarization factor $P(\bm{Q})$ isolates the single-mode transverse correlation function $S_\perp(\bm{Q},E) = \tilde{S}(\bm{Q},E)/P(\bm{Q})$. Because both orthogonal transverse modes contribute equally to the total dynamic spin correlation, $\sum_\alpha S^{\alpha\alpha}(\bm{Q},E) = 2S_\perp(\bm{Q},E) = 2\tilde{S}(\bm{Q},E)/P(\bm{Q})$. By the fluctuation-dissipation theorem, the directionally averaged wavevector-dependent dynamic magnetic susceptibility $\chi''(\bm{Q},E) \equiv \frac{1}{3}\sum_\alpha \chi''_{\alpha\alpha}(\bm{Q},E)$ (in units of $\mu_\mathrm{B}^2$~meV$^{-1}$~f.u.$^{-1}$) is related to the corrected intensity via
\begin{equation}
    \chi''(\bm{Q},E) = \pi g^2 \mu_\mathrm{B}^2 \left(1 - e^{-E / k_\mathrm{B}T}\right) \left[\frac{2}{3}\frac{\tilde{S}(\bm{Q},E)}{P(\bm{Q})}\right].
    \label{eqn:chi_q_e}
\end{equation}

In the hexagonal NiAs-type structure of MnTe ($P6_3/mmc$), the crystallographic unit cell of height $c$ contains two
magnetic Mn sublattices at $\mathbf{d}_1 = (0,0,0)$ and $\mathbf{d}_2 = (0,0,\frac{1}{2})$ separated by $\mathbf{d}_{12} =
\frac{1}{2}\mathbf{c}$. Taking site $0$ on sublattice~$1$ at the origin, the relative site vectors $\mathbf{r}_l$ in
Eq.~(\ref{eqn:spin_corr}) partition into intra-sublattice displacements $\mathbf{r}_l = \mathbf{R}$ (connecting sites on
the same sublattice across crystallographic Bravais lattice vectors $\mathbf{R} = n_1\mathbf{a} + n_2\mathbf{b} +
n_3\mathbf{c}$) and inter-sublattice displacements $\mathbf{r}_l = \mathbf{R} + \mathbf{d}_{12}$ (connecting opposite
sublattices):
\begin{equation}
    S^{\alpha\beta}(\bm{Q},E) = \frac{1}{2\pi\hbar}\int_{-\infty}^{\infty} \mathrm{d}t\, e^{-i\omega t} \left[ \sum_{\mathbf{R}}
    e^{i\bm{Q}\cdot\mathbf{R}} \langle S_{\mathbf{0},1}^\alpha(0) S_{\mathbf{R},1}^\beta(t) \rangle + \sum_{\mathbf{R}}
    e^{i\bm{Q}\cdot(\mathbf{R}+\mathbf{d}_{12})} \langle S_{\mathbf{0},1}^\alpha(0) S_{\mathbf{R},2}^\beta(t) \rangle
    \right].
    \label{eqn:spin_corr_split}
\end{equation}
\noindent
Writing $\bm{Q} = \boldsymbol{\tau} + \bm{Q}$ relative to any crystallographic zone center $\boldsymbol{\tau} =
(H_0, K_0, L_0)$ ($H_0, K_0, L_0 \in \mathbb{Z}$), the Bravais lattice phase satisfies
$e^{i\boldsymbol{\tau}\cdot\mathbf{R}} = 1$, whereas the half-cell sublattice shift contributes a zone-dependent phase
 factor
 \begin{equation}
     e^{i(\boldsymbol{\tau}+\bm{Q})\cdot(\mathbf{R}+\mathbf{d}_{12})} = e^{i\boldsymbol{\tau}\cdot\mathbf{d}_{12}}\,
    e^{i\bm{Q}\cdot(\mathbf{R}+\mathbf{d}_{12})} = e^{i\pi L_0}\, e^{i\bm{Q}\cdot(\mathbf{R}+\mathbf{d}_{12})} =
    (-1)^{L_0}\, e^{i\bm{Q}\cdot(\mathbf{R}+\mathbf{d}_{12})}.
     \label{eqn:sublattice_phase}
 \end{equation}
 \noindent
 Factoring $(-1)^{L_0}$ out of the inter-sublattice sum, the wavevector-dependent dynamic susceptibility in
Eq.~(\ref{eqn:chi_q_e}) separates into reduced-wavevector intra- and inter-sublattice components:
 \begin{equation}
     \chi''(\boldsymbol{\tau}+\bm{Q}, E) = \chi''_\mathrm{intra}(\bm{Q}, E) +
     (-1)^{L_0}\chi''_\mathrm{inter}(\bm{Q}, E).
      \label{eqn:chi_sublattice_decomp}
  \end{equation}
  \noindent
  Averaging over one odd-$L_0$ and one even-$L_0$ primitive crystallographic Brillouin zone $\mathrm{BZ}_\mathrm{cryst}$
($\Delta L = 1$) cancels the $(-1)^{L_0}\chi''_\mathrm{inter}(\bm{Q},E)$ cross-sublattice interference term. This yields the intrinsic single-ion local
dynamic susceptibility,
 \begin{equation}
     \chi''(E) = \frac{1}{2}\left[\chi''_\mathrm{odd}(E) + \chi''_\mathrm{even}(E)\right], \quad
     \chi''_{\mathrm{odd/even}}(E) = \frac{\int_{\mathrm{BZ}_\mathrm{cryst}}
      \chi''_{\mathrm{odd/even}}(\bm{Q},E)\,\mathrm{d}\bm{Q}}{\int_\mathrm{BZ_{cryst}} \mathrm{d}\bm{Q}},
       \label{eqn:chi_e}
   \end{equation}
   \noindent
   from which the squared dynamic magnetic moment per formula unit is obtained by integrating over energy:
\begin{equation}
    M_\mathrm{dyn}^2 = \frac{3}{\pi} \int_{-\infty}^{\infty} \frac{\chi''(E)}{1 - e^{-E / k_\mathrm{B}T}}\, dE.
    \label{eqn:mdyn_int}
\end{equation}

\subsection{Data pre-processing and physical corrections}

This section details the workflow employed to extract the local dynamic susceptibility from the measured inelastic spectra on SEQUOIA (MnTe) and MAPS (Li-MnTe). The starting point for this analysis is a three-dimensional histogram of the entire measured region of reciprocal space at each energy, with energy binning parameters defined by the horizontal error bars in Figure \ref{fig:S_sublattice_sum_rule}. The overall workflow for each of these constant-energy datasets is outlined as follows:

\begin{enumerate}
    \item Artifact removal and background estimation (subtraction) via a three-dimensional punch-and-fill approach,
    \item Magnetic form-factor and polarization factor corrections,
    \item Brillouin zone folding into odd/even reduced zones, and
    \item $D_{6h}$ symmterization of the reduced zones.
\end{enumerate}

\noindent
These steps are described in more detail in the following sections.

\newpage
\subsubsection{Artifact removal and background subtraction}

Because the local moment sum rule requires momentum integration over the full three-dimensional Brillouin zone [Eq.~(\ref{eqn:chi_e})], and because we fold multiple higher-order zones into a single reduced zone for this purpose, an accurate determination and removal of the extrinsic and non-magnetic background across the entire measured region of reciprocal space is necessary to avoid systematic overestimation of the dynamic spectral weight.

We employ an iterative three-dimensional punch-and-fill algorithm for each constant-energy dataset. For visualization purposes we present a representative $(H,K)$ plane slice at $E=20$~meV for both compounds in Figure \ref{fig:S_bkg_subtraction}.
First, spin waves are identified, but not yet removed, by defining an energy-dependent three-dimensional exclusion mask $M_\mathrm{SW}(\bm{Q},E)$ tiled across the crystallographic zone centers $\boldsymbol{\tau} = (H_0, K_0, L_0)$ (Fig. \ref{fig:S_bkg_subtraction}c,h). The mask dimensions at each energy are parameterized from the odd-$L_0$ zones, as these carry finite spin-wave intensity across the entire dispersion. The exclusion volume around each zone center is modeled by an anisotropic ellipsoid,
\begin{equation}
    \left(\frac{q_x}{r_x(E)}\right)^2 + \left(\frac{q_y}{r_y(E)}\right)^2 + \left(\frac{q_z}{r_z(E)}\right)^2 \le 1,
    \label{eqn:sw_mask_ellipsoid}
\end{equation}
\noindent
where $\bm{Q} = \bm{Q} - \boldsymbol{\tau} = (q_x, q_y, q_z)$ is the wavevector displacement relative to the zone center in the orthogonal crystallographic frame $[H, 0, 0]$, $[-0.5K, K, 0]$, and $[0, 0, L]$ (in r.l.u.). In the hexagonal basal plane, rotational symmetry enforces an isotropic spin-wave dispersion in physical momentum space ($q_\perp = \sqrt{Q_x^2 + Q_y^2}$ in $\text{\AA{}}^{-1}$), requiring circular constant-energy contours. Accounting for the reciprocal lattice metric [$Q_{100} = 4\pi / (a\sqrt{3}) \approx 1.7482\,\text{\AA{}}^{-1}$ along $[H, 0, 0]$ and $Q_{0.5, 1, 0} = 2\pi / a = \frac{\sqrt{3}}{2} Q_{100} \approx 1.5140\,\text{\AA{}}^{-1}$ along $[-0.5K, K, 0]$], the in-plane semi-axes satisfy $r_y(E) = \frac{2}{\sqrt{3}} r_x(E)$, which ensures that the mask is strictly circular in the $(H,K)$ planes.

\begin{figure}[H]
\centering
\includegraphics[width=\textwidth]{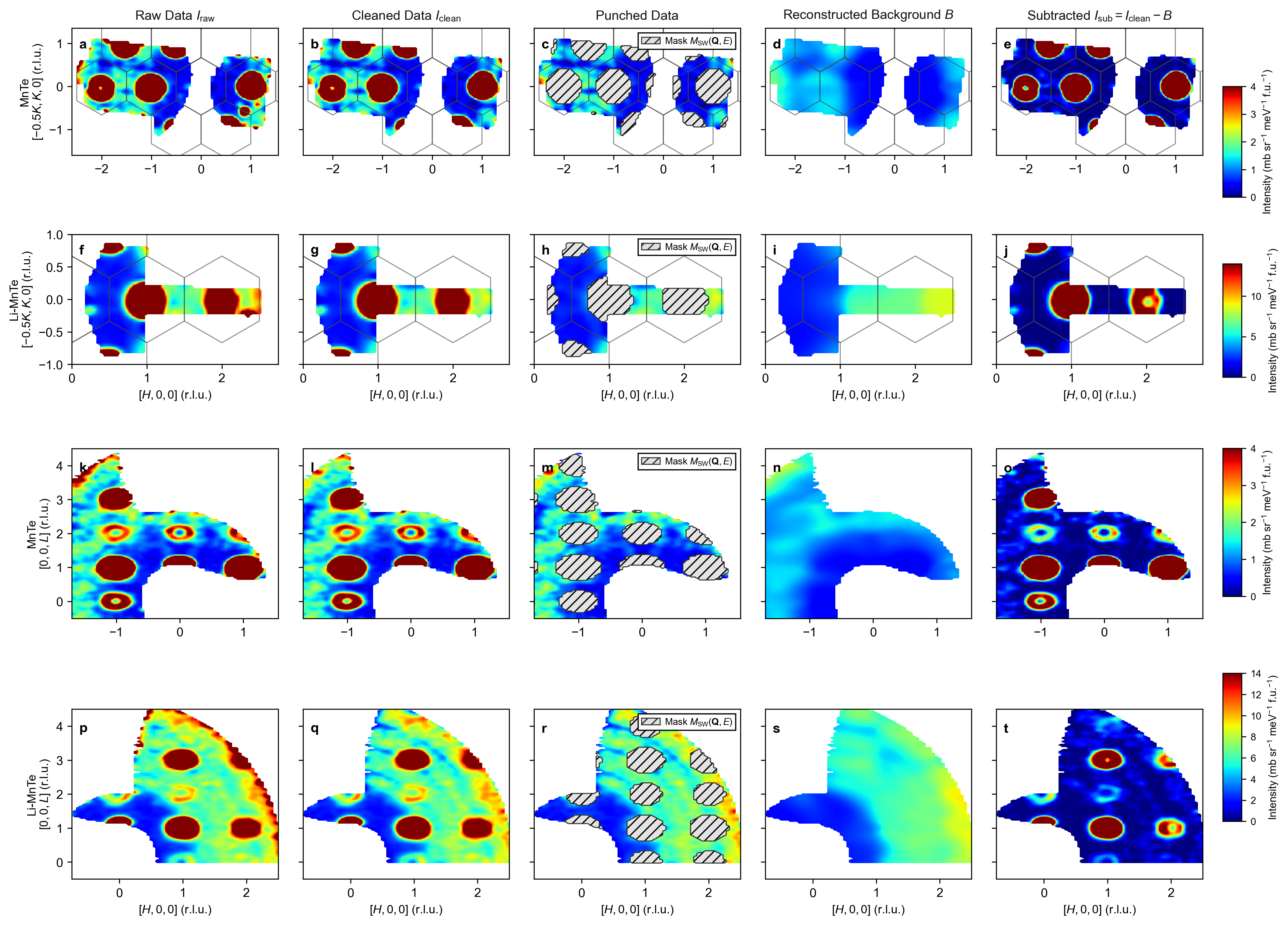}
\caption{Three-dimensional punch-and-fill background subtraction workflow demonstrated at energy transfer $E = 20.0\,\mathrm{meV}$ for MnTe (\textbf{a}--\textbf{e}) and Li-MnTe (\textbf{f}--\textbf{j}) in the $(H,K)$ scattering plane sliced along $[0, 0, L]$ at $L = 1.0 \pm 0.05\,\mathrm{r.l.u.}$ \textbf{a},\textbf{f}, Raw inelastic neutron scattering intensity $I_\mathrm{raw}(\bm{Q},E)$. \textbf{b},\textbf{g}, De-spiked intensity $I_\mathrm{clean}(\bm{Q},E)$ after removing isolated normalization singularities via a $3\times 3\times 3$ median filter. \textbf{c},\textbf{h}, Punched data with the masked spin-wave regions $M_\mathrm{SW}(\bm{Q},E)$ indicated by shaded hatched areas. \textbf{d},\textbf{i}, Reconstructed smooth diffuse and instrumental background $B(\bm{Q},E)$ obtained from normalized Gaussian multi-scale convolution. \textbf{e},\textbf{j}, Background-subtracted magnetic scattering intensity $I_\mathrm{sub}(\bm{Q},E) = I_\mathrm{clean} - B$. Black hexagons delineate the Brillouin zone boundaries in the $(H,K)$ plane.}
\label{fig:S_bkg_subtraction}
\end{figure}

To prevent unphysical single-voxel artifacts from corrupting the background interpolation or the final signal, isolated
intensity spikes arising from solid-angle normalization singularities at detector boundaries are first replaced using a
$3\times 3\times 3$ three-dimensional median filter, yielding the de-spiked data $I_\mathrm{clean}(\bm{Q},E)$
(Fig.~\ref{fig:S_bkg_subtraction}b,g). To then isolate the background from coherent magnetic excitations, voxels inside the
exclusion mask $M_\mathrm{SW}(\bm{Q},E)$ (as well as unmeasured voxels outside the kinematic detector envelope) are
temporarily punched out of the dataset by setting a binary background validity weight $W(\bm{Q},E) = 0$ inside $M_\mathrm{SW}$ and
$W(\bm{Q},E) = 1$ in the remaining background region (Fig.~\ref{fig:S_bkg_subtraction}c,h).
To smoothly interpolate the background across the punched cavities without zero-padding bias (i.e., avoiding artificial
drooping toward zero inside excised regions), we perform a normalized three-dimensional Gaussian convolution:
\begin{equation}
    B_\sigma(\bm{Q},E) = \frac{\left[G_\sigma * \left(W \cdot I_\mathrm{clean}\right)\right](\bm{Q},E)}{\left[G_\sigma * W\right](\bm{Q},E)},
    \label{eqn:norm_gauss_bkg}
\end{equation}
\noindent
where $G_\sigma$ is a three-dimensional Gaussian kernel. To accurately track local background gradients near measured data while robustly bridging across wide exclusion cavities, we employ a smooth multi-scale blending between a primary kernel ($\sigma_1 = 0.15\,\mathrm{r.l.u.}$) and a wider kernel ($\sigma_2 = 0.50\,\mathrm{r.l.u.}$),
\begin{equation}
    B(\bm{Q},E) = \alpha(\bm{Q},E)\, B_{\sigma_1}(\bm{Q},E) + \left[1 - \alpha(\bm{Q},E)\right] B_{\sigma_2}(\bm{Q},E).
    \label{eqn:bkg_multiscale}
\end{equation}
\noindent
The local blending weight $\alpha \in [0, 1]$ is evaluated voxel-by-voxel from the local background weight density $D_1(\bm{Q},E) = [G_{\sigma_1} * W](\bm{Q},E) \in [0, 1]$. In regions with sufficient nearby background support ($D_1 \ge 0.20$), the interpolation relies entirely on the primary kernel ($\alpha = 1$), whereas deep cavity interiors ($D_1 \le 0.02$) transition entirely to the wider kernel ($\alpha = 0$). In the intermediate transition regime ($0.02 < D_1 < 0.20$), the weight varies continuously according to a raised-cosine (Hann-window) function,
\begin{equation}
    \alpha = \frac{1}{2} - \frac{1}{2}\cos\left[\pi \left(\frac{D_1 - 0.02}{0.20 - 0.02}\right)\right],
    \label{eqn:bkg_alpha}
\end{equation}
\noindent
providing smooth blending with zero derivatives at both endpoints to eliminate any gradient steps or seam artifacts in the reconstructed diffuse and instrumental background $B(\bm{Q},E)$ (Fig.~\ref{fig:S_bkg_subtraction}d,i).
The background-subtracted magnetic scattering intensity is then obtained by
\begin{equation}
    I_\mathrm{sub}(\bm{Q},E) = I_\mathrm{clean}(\bm{Q},E) - B(\bm{Q},E),
\end{equation}
\noindent
while voxels outside the kinematic detector envelope remain flagged as unmeasured and are excluded from subsequent multi-zone folding and Brillouin-zone averages (Fig.~\ref{fig:S_bkg_subtraction}e,j).

\subsubsection{Form factor and polarization factor corrections}

\begin{figure*}[t]
    \centering
    \includegraphics[width=\textwidth]{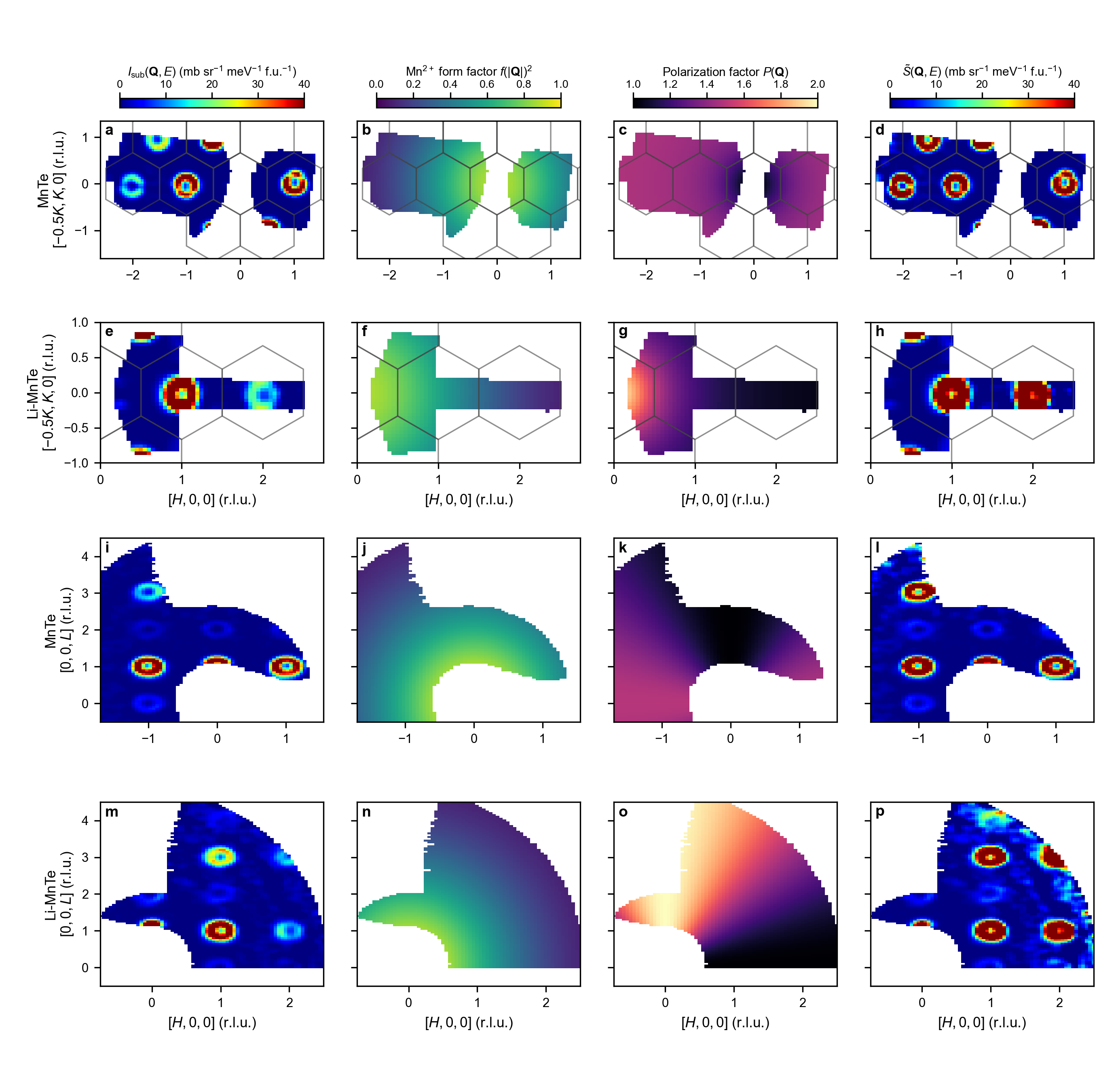}
    \caption{Physical corrections applied to the inelastic neutron scattering data at $E
    = 20.0\,\mathrm{meV}$ in the $(H,K,1)$ scattering plane for MnTe (\textbf{a}--\textbf{d})
    and Li-MnTe (\textbf{e}--\textbf{h}), and in the $(H, 0, L)$ for MnTe (\textbf{i}--\textbf{l}) and Li-MnTe
    (\textbf{m}--\textbf{p}).
     \textbf{a},\textbf{e},\textbf{i},\textbf{m}, Background-subtracted data $I_\mathrm{sub}$
     \textbf{b},\textbf{f},\textbf{j},\textbf{n}, Magnetic form factor $f(|\bm{Q}|)^2$ of Mn$^{2+}$
    [Eq.~(\ref{eqn:form_factor})].
     \textbf{c},\textbf{g},\textbf{k},\textbf{o}, Transverse spin-wave neutron polarization contraction factor $P(\bm{Q})$ for MnTe (\textbf{c},\textbf{k}), and in Li-MnTe (\textbf{g},\textbf{o}). For MnTe, the polarization factor is averaged over the three in-plane magnetic domains.
     \textbf{d},\textbf{h},\textbf{l},\textbf{p}, Single-mode dynamic structure factor after both form factor and polarization corrections}
     \label{fig:S_physical_corrections}
 \end{figure*}

Following background subtraction, physical corrections for the magnetic form factor and polarization factor are applied on a voxel-by-voxel basis to obtain the polarization-contracted dynamic structure factor $\tilde{S}(\bm{Q},E)$ via Eq.~(\ref{eqn:stilde}). The $\mathrm{Mn}^{2+}$ dipole magnetic form factor $f(|\bm{Q}|)$ is calculated using the standard three-exponential analytical parameterization from the International Tables for Crystallography:
\begin{equation}
    f(|\bm{Q}|) = \langle j_0(s) \rangle = \sum_{i=1}^{3} A_i \exp(-a_i s^2) + C,
    \label{eqn:form_factor}
\end{equation}
\noindent
where $s = |\bm{Q}| / (4\pi)$ and the empirical coefficients are $A_1 = 0.4220$, $a_1 = 17.6840$, $A_2 = 0.5948$, $a_2 = 6.0050$, $A_3 = 0.0043$, $a_3 = -0.6090$, and $C = -0.0219$.
The sample-specific transverse spin-wave polarization factor $P(\bm{Q})$ [Eq.~(\ref{eqn:pol_pure}) for MnTe and Eq.~(\ref{eqn:pol_li}) for Li-MnTe] is computed from the local wavevector orientation $\tilde{Q}_z = Q_c / |\bm{Q}|$ for each voxel. Dividing $I_\mathrm{sub}(\bm{Q},E)$ by $f(|\bm{Q}|)^2 P(\bm{Q})$ yields the corrected single-mode structure factor $S_\perp(\bm{Q},E) = \tilde{S}(\bm{Q},E) / P(\bm{Q})$ across multiple Brillouin zones, as illustrated in Fig.~\ref{fig:S_physical_corrections}.

\subsubsection{Sublattice-resolved Brillouin-zone folding}
As established in Eqs.~(\ref{eqn:sublattice_phase}) and (\ref{eqn:chi_sublattice_decomp}), the inter-sublattice interference term reverses sign between odd and even $L_0$ zones via the phase factor $(-1)^{L_0}$, while remaining invariant across all crystallographic Brillouin zones $\boldsymbol{\tau} = (H_0, K_0, L_0)$ that share the same $L_0$ parity. Consequently, the form-factor- and polarization-corrected single-mode structure factor $S_\perp(\bm{Q},E)$ can be folded across higher-order zones of matching parity to maximize reciprocal-space coverage and statistical precision without mixing opposite interference phases. Specifically, all measured Brillouin zones centered at odd $L_0 \in \{1, 3\}$ are folded into an odd-$L_0$ reduced volume $S_{\perp,\mathrm{odd}}(\bm{Q},E)$, whereas all measured zones centered at even $L_0 \in \{0, 2, 4\}$ are folded independently into an even-$L_0$ reduced volume $S_{\perp,\mathrm{even}}(\bm{Q},E)$. Both reduced volumes are evaluated on a uniform grid enclosing the primitive hexagonal prism $\mathrm{BZ}_\mathrm{cryst}$, bounded by $-0.5 \le q_x \le 0.5$, $|q_y| \le \frac{2}{3}(1 - |q_x|)$, and $-0.5 \le q_z \le 0.5$ (in r.l.u.) in the orthogonalized reduced wavevector frame $\bm{Q} = \bm{Q} - \boldsymbol{\tau} = (q_x, q_y, q_z)$.

\begin{figure}[t]
\centering
\includegraphics[width=\textwidth]{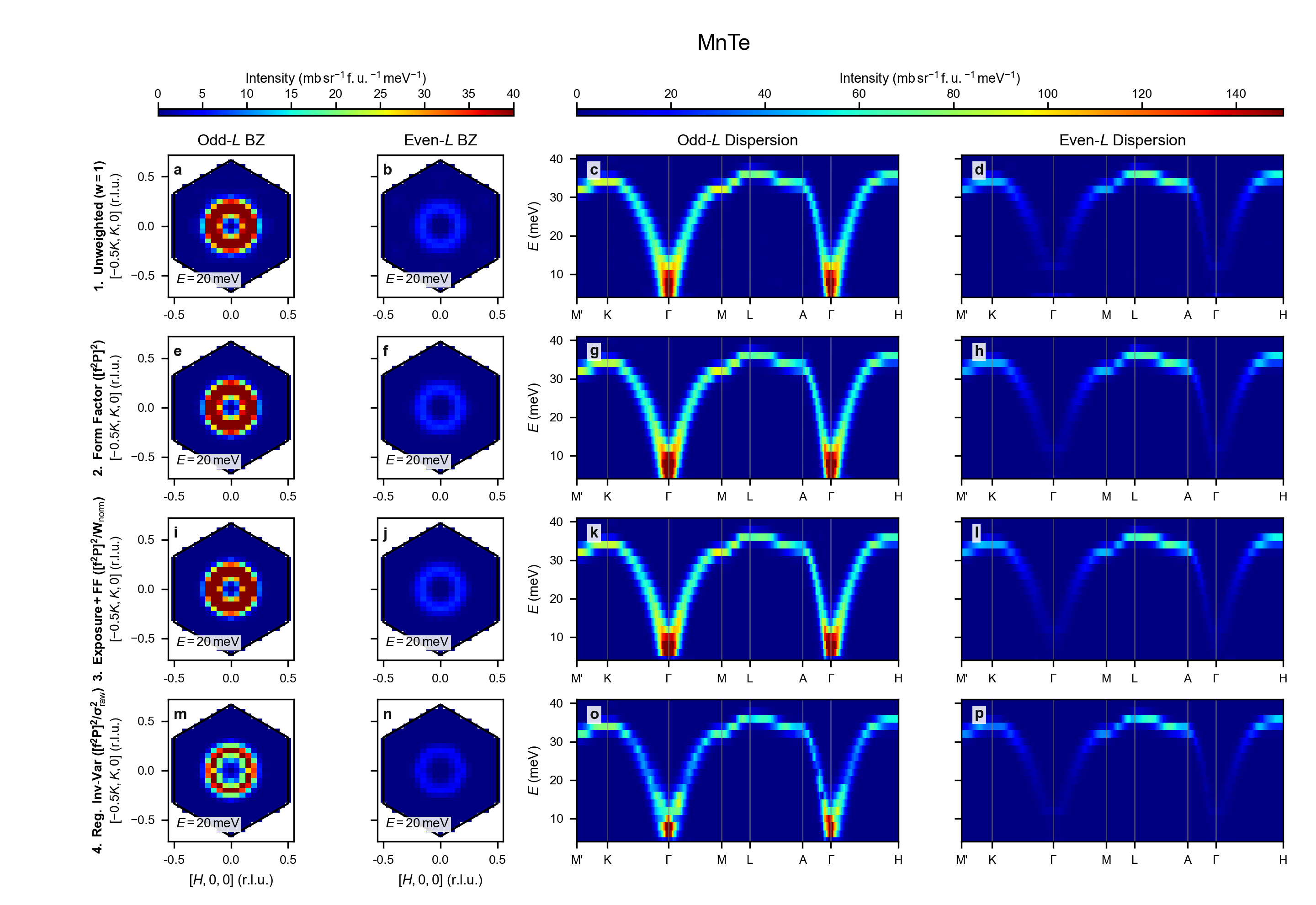}
\caption{Brillouin zone folding for pristine MnTe using four multi-zone weighting schemes: unweighted arithmetic average (\textbf{a}--\textbf{d}), form-factor and multi-domain polarization weighting (\textbf{e}--\textbf{h}), exposure-normalized form-factor weighting (\textbf{i}--\textbf{l}), and regularized inverse-variance weighting (\textbf{m}--\textbf{p}). Panels \textbf{a,e,i,m} depict constant-energy slices of the $(H,K)$ plane of the reduced zones at $E = 20.0\,\mathrm{meV}$ for odd $L$ and (\textbf{b,f,j,n}) even $L$. Panels \textbf{c,g,k,o} depict the dispersion for the odd-$L$ and (\textbf{d,h,l,p}) even-$L$ reduced zones. The coarse energy binning shown here reflects that used throughout the total moment sum rule workflow.}
\label{fig:S_bz_folding_mnte}
\end{figure}

\begin{figure}[t]
\centering
\includegraphics[width=\textwidth]{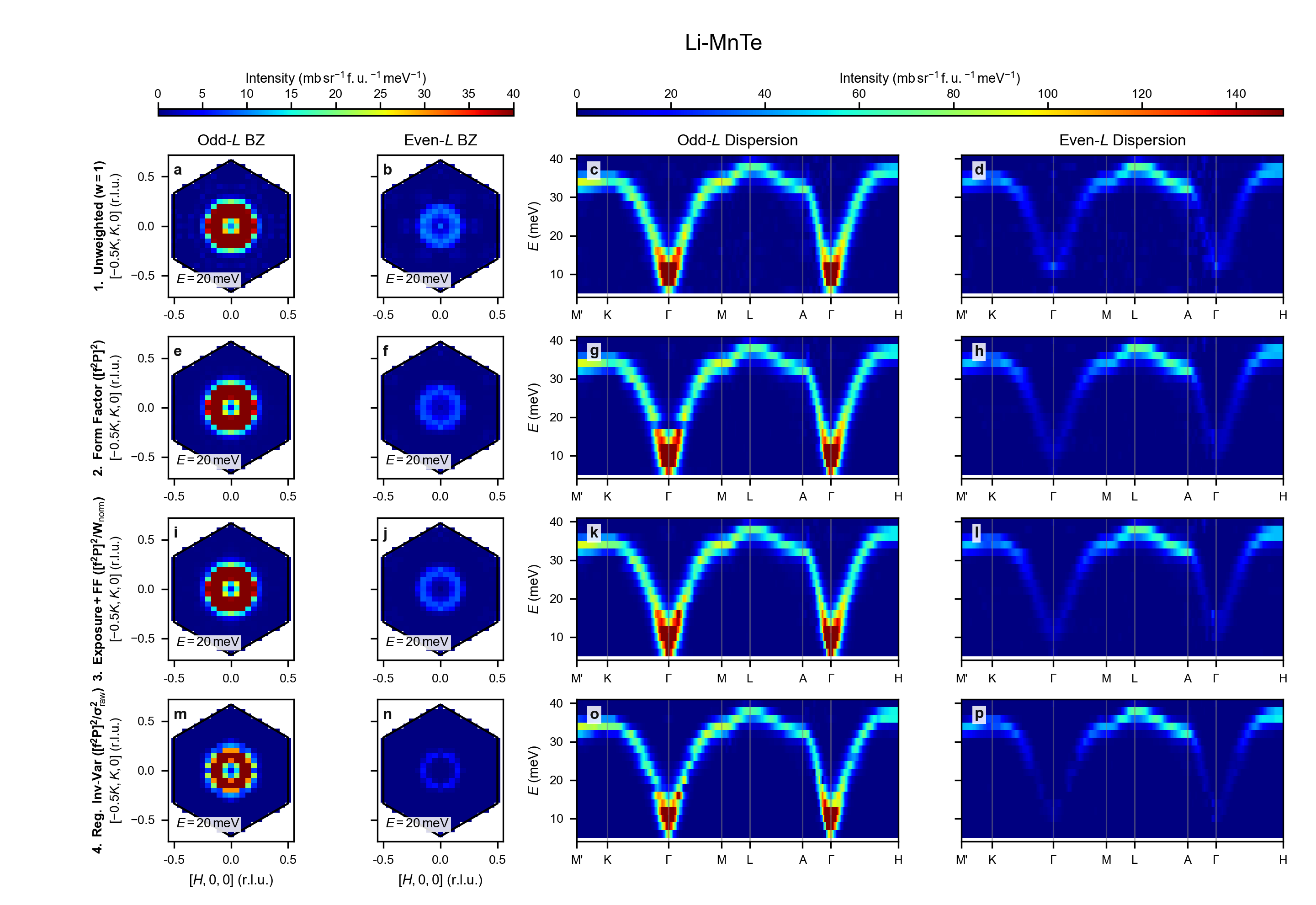}
\caption{Brillouin zone folding for Li-MnTe using four multi-zone weighting schemes: unweighted arithmetic average (\textbf{a}--\textbf{d}), form-factor and multi-domain polarization weighting (\textbf{e}--\textbf{h}), exposure-normalized form-factor weighting (\textbf{i}--\textbf{l}), and regularized inverse-variance weighting (\textbf{m}--\textbf{p}). Panels \textbf{a,e,i,m} depict constant-energy slices of the $(H,K)$ plane of the reduced zones at $E = 20.0\,\mathrm{meV}$ for odd $L$ and (\textbf{b,f,j,n}) even $L$. Panels \textbf{c,g,k,o} depict the dispersion for the odd-$L$ and (\textbf{d,h,l,p}) even-$L$ reduced zones. The coarse energy binning shown here reflects that used throughout the total moment sum rule workflow.}
\label{fig:S_bz_folding_limnte}
\end{figure}

In order to accomplish this, we evaluate four different voxel weighting schemes and their impact on the total moment sum rule. When summing $S_{\perp}(\mathbf{\tau}+\bm{Q},E)$ from different zones of the same parity to obtain the odd/even reduced zone structure factors [$S_{\perp,\mathrm{odd/even}}(\bm{Q}, E)$], each weighting scheme is applied via
\begin{equation}
    S_{\perp,\mathrm{odd/even}}(\bm{Q}, E) = \frac{\sum_{\boldsymbol{\tau}} w(\boldsymbol{\tau} + \bm{Q}, E)\, S_\perp(\boldsymbol{\tau} + \bm{Q}, E)}{\sum_{\boldsymbol{\tau}} w(\boldsymbol{\tau} + \bm{Q}, E)},
    \label{eqn:weighted_bz_fold}
\end{equation}
\noindent
where $w(\mathbf{\tau}+\bm{Q},E)$ is the weight applied to each voxel.

As a benchmark, we first evaluate the reduced zone with a uniform weight across the entire measured region of reciprocal space:
\begin{equation}
    w_1(\mathbf{\tau}+\bm{Q},E)=1.
\end{equation}
 The results from this weighting scheme are depicted in Figs. \ref{fig:S_bz_folding_mnte}a-d and \ref{fig:S_bz_folding_limnte}a-d. While strictly unbiased and apparently acceptable when applied to the SEQUOIA data on pristine MnTe, it introduces significant random noise when applied to the MAPS data collected on Li-MnTe. This is likely due to noisy data at higher $|\bm{Q}|$ where the form factor is weak, leading to amplified noise after applying the form-factor correction.
This result suggests that a non-uniform weighting scheme which places importance on higher-quality (lower-$\bm{Q}$) zones is warranted here.

There are several options to choose from in this regard. First, one can weight by the magnetic form factor and polarization factor, since these are the dominant terms dictating which zones are expected to have comparatively better statistics. This approach is also unbiased by the actual measured data, but identically does not explicitly take into account the data quality on a voxel-by-voxel basis. This weighting scheme is given by
\begin{equation}
    w_2(\mathbf{\tau}+\bm{Q},E)=[f(|\bm{Q}|)^2P(\bm{Q})]^2.
\end{equation}
Alternatively, one can additionally scale by the ratio $\frac{1}{W_\mathrm{norm}}=\frac{I_\mathrm{raw}}{\sigma^2_\mathrm{raw}}$. Since $I_\mathrm{raw} = W_\mathrm{norm} N$ and $\sigma_\mathrm{raw}^2 = W_\mathrm{norm}^2 N$ for normalized Poisson counts, the scattered count $N$ cancels in the
ratio $\frac{1}{W_\mathrm{norm}} = \frac{I_\mathrm{raw}}{\sigma_\mathrm{raw}^2}$, isolating an unbiased effective detector-exposure term. This weight is given by
\begin{equation}
    w_3(\mathbf{\tau}+\bm{Q},E)=\frac{[f(|\bm{Q}|)^2P(\bm{Q})]^2}{W_\mathrm{norm}(\bm{Q},E)},
\end{equation}
though one must filter appropriately for edge cases such as low-count or empty voxels which can result in noisy weights within this scheme.
Lastly, one may attempt to weight by the inverse of the propagated variance such that
\begin{equation}
    w_4(\boldsymbol{\tau}+\bm{Q},E) = \frac{[f(|\bm{Q}|)^2 P(\bm{Q})]^2}{\sigma^2_\mathrm{raw}(\bm{Q},E)},
\end{equation}
with $\sigma_\mathrm{raw}^2$ floored at low counts to prevent division by zero in empty voxels. However, because $\sigma_\mathrm{raw}^2 \propto N$ where $N$ is the number of counts at the detector, this approach systematically penalizes voxels with large counts (i.e., the spin wave signal), resulting in an underestimate of the total spectral weight (this is verified in Table \ref{tab:sumrule_summary}).
After the chosen weighting scheme is applied, the symmetry operations of the $D_{6h}$ point group are applied to each reduced zone to yield a symmetrized Brillouin zone as shown in Figures \ref{fig:S_bz_folding_mnte} and \ref{fig:S_bz_folding_limnte}.

In the unweighted scheme, excess intensity is evident away from the spin waves (e.g., at the origin of Fig. \ref{fig:S_bz_folding_limnte}b or at the $\Gamma$ point in Fig. \ref{fig:S_bz_folding_limnte}d).
Weighting by the form factor and polarization factor alone (Fig. \ref{fig:S_bz_folding_limnte}e-h) or directly by the inverse variance (Fig. \ref{fig:S_bz_folding_limnte}m-p) lead to abrupt dips in the spectral weight and jagged edges along the dispersion, likely due to unattenuated weights on low-count detector edges in $w_2$ and excessive weights on low-count voxels (coupled with peak suppression) in $w_4$.
Visual inspection of the results demonstrates that combined weighting by the exposure, form factor, and polarization factor together yield a satisfactory recreation of the spin wave intensity across the Brillouin zone without substantial gaps or noise.

\subsection{Energy integration and total moment evaluation}
The local dynamic susceptibility $\chi''(E)$ is computed by numerical integration of Eq.~(\ref{eqn:chi_q_e}) over the primitive magnetic Brillouin zone:
\begin{equation}
    \chi''(E) = \frac{1}{V_\mathrm{BZ}}\int_\mathrm{BZ} \chi''(\bm{Q},E)\, d\bm{Q} \approx \frac{1}{N_\mathrm{vox}} \sum_{\bm{Q} \in \mathrm{BZ}} \chi''(\bm{Q},E),
    \label{eqn:chi_e_discrete}
\end{equation}
\noindent
where $V_\mathrm{BZ}$ is the volume of the Brillouin zone and  $N_\mathrm{vox}$ is the total number of discrete voxels enclosed within the Brillouin zone. The integrated squared dynamic magnetic moment per formula unit is then obtained by numerical integration over energy transfer:
\begin{equation}
    M_\mathrm{dyn}^2 = \frac{3}{\pi} \int_{E_\mathrm{min}}^{E_\mathrm{max}} \frac{\chi''(E)}{1 - \exp(-E/k_\mathrm{B}T)}\, dE,
    \label{eqn:cum_mdyn}
\end{equation}
\noindent
where the lower cutoff $E_\mathrm{min}$ corresponds to the low-energy boundary of the measurement ($4.5\,\mathrm{meV}$ for MnTe and $5.0\,\mathrm{meV}$ for Li-MnTe) and the upper integration boundary is $E_\mathrm{max} = 40.0\,\mathrm{meV}$, capturing approximately the full spin-wave bandwidth. For MnTe, because the spin waves are nearly gapless within the resolution of our measurements, we acknowledge that this conservative low-energy cutoff excludes some of the lowest-energy dynamic spin spectral weight which overlaps with the strong elastic signal, thus potentially leading to a slight underestimation of the dynamic moment.

\begin{figure}[t]
\centering
\includegraphics[width=\textwidth]{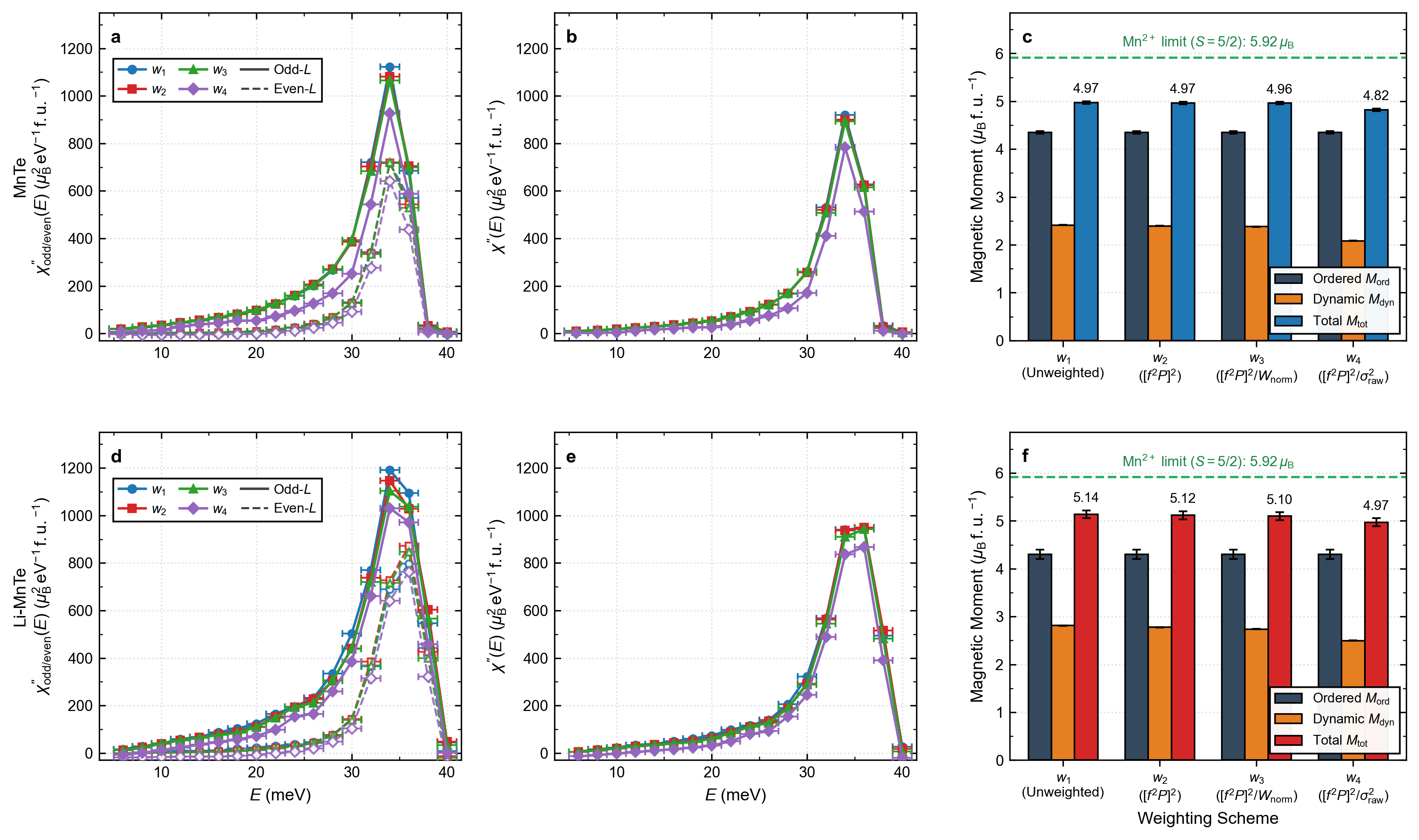}
\caption{Comparison of total moment sum rule results for different zone-folding weighting schemes ($w_1\text{--}w_4$) for pristine MnTe (\textbf{a}--\textbf{c}) and Li-MnTe (\textbf{d}--\textbf{f}). \textbf{a},\textbf{d} Momentum-integrated dynamic magnetic susceptibility $\chi''_\mathrm{odd}(E)$ for the odd-$L$ reduced zone (solid curves, filled markers) and $\chi''_\mathrm{even}(E)$ for the even-$L$ reduced zone (dashed curves, open markers). \textbf{b},\textbf{e} Sublattice-averaged local dynamic susceptibility $\chi''(E) = \frac{1}{2}[\chi''_\mathrm{odd}(E) + \chi''_\mathrm{even}(E)]$. Horizontal error bars denote energy-integration ranges. Vertical error bars are smaller than the width of the data markers, and represent the 1$\sigma$ uncertainties after weighted zone folding. \textbf{c},\textbf{f} Comparison of the ordered moment $M_\mathrm{ord}$, dynamic moment $M_\mathrm{dyn}$, and total local moment $M_\mathrm{tot} = \sqrt{M_\mathrm{ord}^2 + M_\mathrm{dyn}^2}$ across all four weighting schemes. Horizontal dashed green lines denote the theoretical high-spin $\mathrm{Mn}^{2+}$ limit ($g\sqrt{S(S+1)} \approx 5.9\,\mu_\mathrm{B}$ for $S = 5/2$).}
\label{fig:S_sublattice_sum_rule}
\end{figure}

Figure~\ref{fig:S_sublattice_sum_rule} and Table~\ref{tab:sumrule_summary} summarize the momentum-integrated dynamic susceptibility $\chi''(E)$ and the resulting dynamic moment across all four zone-folding weighting schemes ($w_1\text{--}w_4$). In agreement with Eq.~(\ref{eqn:chi_sublattice_decomp}), integrating exclusively over the odd-$L_0$ reduced zone (solid curves in Fig.~\ref{fig:S_sublattice_sum_rule}a,d) overshoots the true single-ion spectral weight due to constructive inter-sublattice interference near $\Gamma$, whereas integrating exclusively over the even-$L_0$ reduced zone (dashed curves in Fig.~\ref{fig:S_sublattice_sum_rule}a,d) undershoots the spectral weight due to destructive sublattice interference. Taking the equal sublattice average $\chi''(E) = \frac{1}{2}[\chi''_\mathrm{odd}(E) + \chi''_\mathrm{even}(E)]$ (Fig.~\ref{fig:S_sublattice_sum_rule}b,e) eliminates the $(-1)^{L_0}\chi''_\mathrm{inter}(\bm{Q}, E)$ cross-sublattice interference term and isolates the intrinsic single-ion dynamic response.

\begin{table}[t]
\centering
\caption{Summary of integrated dynamic moments, ordered moments from neutron diffraction, total local moments $M_\mathrm{tot} = \sqrt{M_\mathrm{ord}^2 + M_\mathrm{dyn}^2}$, and percentage recovery of the theoretical high-spin $\mathrm{Mn}^{2+}$ limit [$g\sqrt{S(S+1)} \approx 5.9\,\mu_\mathrm{B}$ for $S = 5/2$] across the four zone-folding weighting schemes for MnTe and Li-MnTe.}
\label{tab:sumrule_summary}
\begin{tabular}{|l|c|c|c|c|}
\hline
Quantity & $w_1$ & $w_2$ & $w_3$ & $w_4$ \\ \hline
\multicolumn{5}{|c|}{\textbf{MnTe} (SEQUOIA, $T = 5\,\mathrm{K}$)} \\ \hline
$M^2_\mathrm{dyn,odd}$ ($\mu_\mathrm{B}^2$) & 7.87(2) & 7.81(2) & 7.74(2) & 5.79(2) \\ \hline
$M_\mathrm{dyn,even}^2$ ($\mu_\mathrm{B}^2$) & 3.77(2) & 3.66(2) & 3.61(2) & 2.87(2) \\ \hline
$M_\mathrm{dyn}^2$ ($\mu_\mathrm{B}^2$) & 5.82(2) & 5.74(2) & 5.68(2) & 4.33(2) \\ \hline
Dynamic Moment $M_\mathrm{dyn}$ ($\mu_\mathrm{B}$) & 2.41(1) & 2.39(1) & 2.38(1) & 2.08(1) \\ \hline
Ordered Moment $M_\mathrm{ord}$ ($\mu_\mathrm{B}$) & 4.35(3) & 4.35(3) & 4.35(3) & 4.35(3) \\ \hline
Total Moment $M_\mathrm{tot}$ ($\mu_\mathrm{B}$) & 4.97(3) & 4.97(3) & 4.96(3) & 4.82(3) \\ \hline
Sum Rule Recovery (\%) & 84.1\% & 83.9\% & 83.8\% & 81.5\% \\ \hline
\multicolumn{5}{|c|}{\textbf{Li-MnTe} (MAPS, $T = 5\,\mathrm{K}$)} \\ \hline
$\langle m_\mathrm{dyn}^2 \rangle_\mathrm{odd}$ ($\mu_\mathrm{B}^2$) & 10.64(2) & 10.16(2) & 9.93(2) & 8.46(2) \\ \hline
$\langle m_\mathrm{dyn}^2 \rangle_\mathrm{even}$ ($\mu_\mathrm{B}^2$) & 5.17(2) & 5.26(2) & 5.07(2) & 3.99(2) \\ \hline
$\langle m_\mathrm{dyn}^2 \rangle_\mathrm{sub}$ ($\mu_\mathrm{B}^2$) & 7.91(2) & 7.71(2) & 7.50(2) & 6.23(2) \\ \hline
Dynamic Moment $M_\mathrm{dyn}$ ($\mu_\mathrm{B}$) & 2.81(1) & 2.78(1) & 2.74(1) & 2.50(1) \\ \hline
Ordered Moment $M_\mathrm{ord}$ ($\mu_\mathrm{B}$) & 4.30(10) & 4.30(10) & 4.30(10) & 4.30(10) \\ \hline
Total Moment $M_\mathrm{tot}$ ($\mu_\mathrm{B}$) & 5.14(8) & 5.12(8) & 5.10(8) & 4.97(9) \\ \hline
Sum Rule Recovery (\%) & 86.8\% & 86.5\% & 86.2\% & 84.0\% \\ \hline
\end{tabular}
\end{table}

Across the three count-independent, statistically unbiased weighting schemes ($w_1, w_2, w_3$), both the energy-dependent susceptibility profiles $\chi''(E)$ and the integrated dynamic moments appear consistent, yielding $M_\mathrm{dyn}^2 = 5.68(2)\text{ to }5.82(2)\,\mu_\mathrm{B}^2$ [$M_\mathrm{dyn} = 2.38(1)\text{ to }2.41(1)\,\mu_\mathrm{B}$] for MnTe and $M_\mathrm{dyn}^2 = 7.50(2)\text{ to }7.91(2)\,\mu_\mathrm{B}^2$ [$M_\mathrm{dyn} = 2.74(1)\text{ to }2.81(1)\,\mu_\mathrm{B}$] for Li-MnTe. Combining these dynamic moments in quadrature with the static ordered moments determined from neutron diffraction [$M_\mathrm{ord} = 4.35(3)\,\mu_\mathrm{B}$ for MnTe and $4.30(10)\,\mu_\mathrm{B}$ for Li-MnTe] via Eq.~(\ref{eqn:sumrule_def}) gives total local moments of $M_\mathrm{tot} = 4.96(3)\text{ to }4.97(3)\,\mu_\mathrm{B}$ ($83.8\text{ to }84.1\%$ of the theoretical $S = 5/2$ limit) for MnTe and $M_\mathrm{tot} = 5.10(8)\text{ to }5.14(8)\,\mu_\mathrm{B}$ ($86.2\text{ to }86.8\%$) for Li-MnTe (Fig.~\ref{fig:S_sublattice_sum_rule}c,f and Table~\ref{tab:sumrule_summary}). The exposure-normalized form-factor weighting scheme ($w_3$), which suppresses both high-$|\bm{Q}|$ form-factor noise amplification and low-exposure detector boundary artifacts without count-rate bias, yields $M_\mathrm{tot} = 4.96(3)\,\mu_\mathrm{B}$ ($83.8\%$) for MnTe and $M_\mathrm{tot} = 5.10(8)\,\mu_\mathrm{B}$ ($86.2\%$) for Li-MnTe, and these values are chosen as representative results to present in the main text. By contrast, weighting directly by the inverse empirical variance ($w_4$) systematically depresses $\chi''(E)$ near the spin-wave peak maxima where $\sigma_\mathrm{raw}^2 \propto N$ is largest, underestimating $M_\mathrm{dyn}^2$ by $24\%$ in MnTe [$4.33(2)\,\mu_\mathrm{B}^2$] and $17\%$ in Li-MnTe [$6.23(2)\,\mu_\mathrm{B}^2$] relative to $w_3$ and reducing $M_\mathrm{tot}$ to $4.82(3)\,\mu_\mathrm{B}$ ($81.5\%$) and $4.97(9)\,\mu_\mathrm{B}$ ($84.0\%$), respectively.

For all unbiased weighting  ($w_1, w_2, w_3$), the minor unrecovered spectral weight ($\sim 13\text{--}16\%$) is expected within the precision of the technique. Crucially, the total local magnetic moments $M_\mathrm{tot}$ of pristine MnTe and hole-doped Li-MnTe agree within experimental uncertainty across every weighting scheme [$4.96(3)\,\mu_\mathrm{B}$ vs.\ $5.10(8)\,\mu_\mathrm{B}$ for $w=w_3$], demonstrating that Li doping preserves the local $\mathrm{Mn}^{2+}$ moment while transferring spectral weight from the static ordered channel into enhanced dynamic spin fluctuations.
}

\newpage

\section{Momentum- and energy-integration ranges for the data}

In Figures 2, S2, and S3, the momentum integration directions and the plotting direction (if applicable) are mutually perpendicular. The integration ranges were chosen to ensure that the integrated volume in reciprocal space (in units of \AA{}$^{-1}$) is consistent across different directions throughout each figure. Integration ranges for other figures are provided in Table \ref{range2}. For datasets where altermagnetic splitting is expected, such small integration ranges may be necessary to resolve the splitting clearly.

\begin{table}[!ht]
\centering
\caption{Momentum-integration ranges for Fig. 2 (a)(b), Fig. S2, and S3}
\label{range1}
\begin{tabular}{|c|c|c|}
\hline
Path & \begin{tabular}[c]{@{}c@{}}Momentum 1 (r.l.u.)\end{tabular} & \begin{tabular}[c]{@{}c@{}}Momentum 2 (r.l.u.)\end{tabular} \\ \hline
$\Gamma$-M        & {[}$-0.5K$, $K$, 0{]}, $\pm$0.0578 & {[}0, 0, $L${]}, $\pm$0.093 \\ \hline
M-L            & {[}$H$, 0, 0{]}, $\pm$0.05 & {[}0, 0, $L${]}, $\pm$0.093 \\ \hline
L-A            & {[}$-0.5K$, $K$, 0{]}, $\pm$0.0578 & {[}0, 0, $L${]}, $\pm$0.093 \\ \hline
$\Gamma$-A        & {[}$H$, 0, 0{]}, $\pm$0.05 & {[}0, 0, $L${]}, $\pm$0.093 \\ \hline
$\Gamma$-H        & {[}$K$, $-K$, 0{]}, $\pm$0.05 & {[}$L$, $L$, $-6.89L${]}, $\pm$0.012 \\ \hline
$\Gamma$-K-M$^{'}$     & {[}$K$, $-K$, 0{]}, $\pm$0.05 & {[}0, 0, $L${]}, $\pm$0.093 \\ \hline
\end{tabular}
\end{table}

\begin{table}[H]
\centering
\caption{Momentum-integration ranges for the folded reduced-zone dispersions in Fig.~S9 and Fig.~S10}
\label{range3}
\begin{tabular}{|c|c|c|}
\hline
Path & \begin{tabular}[c]{@{}c@{}}Momentum 1 (r.l.u.)\end{tabular} & \begin{tabular}[c]{@{}c@{}}Momentum 2 (r.l.u.)\end{tabular} \\ \hline
$M'\text{--}K$      & {[}$H$, 0, 0{]}, $\pm$0.025         & {[}0, 0, $L${]}, $\pm$0.025 \\ \hline
$K\text{--}\Gamma$  & {[}$-0.5K$, $K$, 0{]}, $\pm$0.025   & {[}0, 0, $L${]}, $\pm$0.025 \\ \hline
$\Gamma\text{--}M$  & {[}$-0.5K$, $K$, 0{]}, $\pm$0.025   & {[}0, 0, $L${]}, $\pm$0.025 \\ \hline
$M\text{--}L$       & {[}$H$, 0, 0{]}, $\pm$0.025         & {[}$-0.5K$, $K$, 0{]}, $\pm$0.025 \\ \hline
$L\text{--}A$       & {[}$-0.5K$, $K$, 0{]}, $\pm$0.025   & {[}0, 0, $L${]}, $\pm$0.025 \\ \hline
$A\text{--}\Gamma$  & {[}$H$, 0, 0{]}, $\pm$0.025         & {[}$-0.5K$, $K$, 0{]}, $\pm$0.025 \\ \hline
$\Gamma\text{--}H$  & {[}$K$, $-K$, 0{]}, $\pm$0.025      & {[}$L$, $L$, $-6.89L${]}, $\pm$0.025 \\ \hline
\end{tabular}
\end{table}

\begin{table}[H]
\centering
\caption{Momentum-integration ranges for Fig.~2(c)-(j)}
\label{range4}
\begin{tabular}{|c|c|c|c|}
\hline
Path & \begin{tabular}[c]{@{}c@{}}Momentum 1 (r.l.u.)\end{tabular} & \begin{tabular}[c]{@{}c@{}}Momentum 2 (r.l.u.)\end{tabular} & \begin{tabular}[c]{@{}c@{}}Momentum 3 (r.l.u.)\end{tabular}  \\ \hline
$K$-$\Gamma$     & {[}$-K$, $0.5K$, 0{]}, $\pm$0.04
                 & {[}0, $K$, 0{]}, $\pm$0.03
                 & {[}0, 0, $L${]}, $\pm$0.06 \\ \hline
$\Gamma$-$M$     & {[}$H$, 0, 0{]}, $\pm$0.03
                 & {[}$-0.5K$, $K$, 0{]}, $\pm$0.03
                 & {[}0, 0, $L${]}, $\pm$0.06 \\ \hline
$M$-$L$          & {[}0, 0, $L${]}, $\pm$0.06
                 & {[}$-K$, $0.5K$, 0{]}, $\pm$0.03
                 & {[}0, $K$, 0{]}, $\pm$0.03 \\ \hline
$L$-$A$          & {[}$H$, 0, 0{]}, $\pm$0.03
                 & {[}$-0.5K$, $K$, 0{]}, $\pm$0.03
                 & {[}0, 0, $L${]}, $\pm$0.06 \\ \hline
$\Gamma$-$A$     & {[}0, 0, $L${]}, $\pm$0.06
                 & {[}$-K$, $0.5K$, 0{]}, $\pm$0.03
                 & {[}0, $K$, 0{]}, $\pm$0.03 \\ \hline
$\Gamma$-$H$     & {[}$K$, $-0.5K$, 0{]}, $\pm$0.04
                 & {[}0, $K$, 0{]}, $\pm$0.03
                 & {[}$2L$, $-L$, $6.99L${]}, $\pm$0.009 \\ \hline
\end{tabular}
\end{table}

\begin{table}[H]
\centering
\caption{Momentum- and energy-integration ranges for other figures}
\label{range2}
\begin{tabular}{|c|c|c|c|c|}
\hline
Data & Momentum  1 (r.l.u.)  & Momentum  2 (r.l.u.)      & Momentum  3 (r.l.u.)  & Energy  (meV) \\ \hline

Fig. 3(a)(b) & {[}$H$, 0, 0{]}, $\pm$0.03 & {[}$-0.5K$, $K$, 0{]}, $\pm$0.02 & N/A                   & N/A           \\ \hline
Fig. 3(c)(d) & {[}$-0.5K$, $K$, 0{]}, $\pm$0.08 & {[}$L$, 0, $-3.45 L${]}, $\pm$0.01 & N/A                   & N/A           \\ \hline
Fig. 3(f) & {[}$H$, 0, 0{]}, $\pm$0.02 & {[}$-0.5K$, $K$, 0{]}, $\pm$0.04 & {[}0, 0, $0.5L${]}, $\pm$0.04 & N/A           \\ \hline
Fig. 3(g) & {[}$0.5H$, 0, $0.5H${]}, $\pm$0.04 & {[}$-0.5K$, $K$, 0{]}, $\pm$0.04 & {[}$0.29L$, 0, $-L${]}, $\pm$0.02 & N/A           \\ \hline
Fig. 4(a) & {[}0, 0, $L${]}, $\pm$0.02 & N/A                       & N/A                   & $\pm$0.5         \\ \hline
Fig. S4(a) & {[}$H$, 0, 0{]}, $\pm$0.02 & {[}$-0.5K$, $K$, 0{]}, $\pm$0.04                       & N/A                   & N/A         \\ \hline
Fig. S4(b)& {[}$-0.5K$, $K$, 0{]}, $\pm$0.04 & {[}$L$, 0, $-3.45 L${]}, $\pm$0.01 & N/A                   & N/A         \\ \hline
\begin{tabular}[c]{@{}c@{}}Fig. S7(a)-(j)\\S8(a)-(h)\end{tabular} & {[}0, 0, $L${]}, $\pm$0.05 & N/A & N/A & 20.0, $\pm$1.0 \\ \hline
\begin{tabular}[c]{@{}c@{}}Fig. S7(k)-(t)\\S8(i)-(p)\end{tabular}  & {[}$-0.5K$, $K$, 0{]}, $\pm$0.05 & N/A & N/A & 20.0, $\pm$1.0 \\ \hline
\begin{tabular}[c]{@{}c@{}}Fig. S9, S10\\ (cols 1--2)\end{tabular} & {[}0, 0, $q_z${]}, $\pm$0.025 & N/A & N/A & 20.0, $\pm$1.0 \\ \hline
\end{tabular}
\end{table}

\end{document}